\documentclass[fleqn,usenatbib]{mnras}

\usepackage{newtxtext,newtxmath}

\usepackage{array}

\usepackage{lineno}

\usepackage[T1]{fontenc}

\DeclareRobustCommand{\VAN}[3]{#2}
\let\VANthebibliography\thebibliography
\def\thebibliography{\DeclareRobustCommand{\VAN}[3]{##3}\VANthebibliography}

\usepackage{graphicx}	
\usepackage{amsmath}	
\usepackage{amssymb}	

\title[Improving mass modelling through a perturbative approach]{Improving parametric mass modelling of lensing clusters through a perturbative approach}

\author[Beauchesne et al.]{
Benjamin Beauchesne,$^{1,2}$
\thanks{E-mail: benjamin.beauchesne@epfl.ch}
Benjamin Cl\'ement,$^{1}$
Johan Richard,$^{2}$
and
Jean-Paul Kneib$^{1,3}$
\\
$^{1}$Institute of Physics, Laboratory of Astrophysics, Ecole Polytechnique Fédérale de Lausanne (EPFL), Observatoire de Sauverny, 1290 Versoix, Switzerland\\
$^{2}$Univ Lyon, Univ Lyon1, Ens de Lyon, CNRS, Centre de Recherche Astrophysique de Lyon UMR5574, 69230, Saint-Genis-Laval, France\\
$^{3}$Aix Marseille Université, CNRS, LAM (Laboratoire d'Astrophysique de Marseille) UMR 7326, 13388, Marseille, France
}

\date{Accepted XXX. Received YYY; in original form ZZZ}

\pubyear{2020}

\begin{document}
\setpagewiselinenumbers
\label{firstpage}
\pagerange{\pageref{firstpage}--\pageref{lastpage}}
\maketitle

\begin{abstract}
We present a new method to model the mass distribution of galaxy clusters that combines a \textit{parametric} and a \textit{free-form} approach to reconstruct cluster cores with strong lensing constraints. It aims at combining the advantages of both approaches, by keeping the robustness of the \textit{parametric} component with an increased flexibility thanks to a \textit{free-form} surface of B-spline functions. We demonstrate the capabilities of this new approach on the simulated cluster \textit{Hera}, which has been used to evaluate lensing codes for the analysis of the Frontier Fields clusters. The method leads to better reproduction of the constraints, with an improvement by a factor $\sim3-4$ on the root-mean-square error on multiple-image positions, when compared to parametric-only approaches. The resulting models show a better accuracy in the reconstruction of the amplitude of the convergence field while conserving a high fidelity on other lensing observables already well reproduced. We make this method publicly available through its implementation in the \textit{Lenstool} software.
\end{abstract}




\section{Introduction}
Galaxy clusters are among the biggest structures that are gravitationally bound in the Universe, and they are located at the intersection of the filamentary structures that form the cosmic web. Hierarchical models of galaxy evolution and the standard $\Lambda$CDM cosmological model predict that clusters aggregate matter at the intersections of the cosmological filaments. This hierarchical growth, in particular through merging events, is continuously increasing cluster masses.

The study of galaxy clusters through high-resolution space-based observations have significantly contributed to the study of the importance of the Dark Matter (DM) content \citep{Natarajan2002,Limousin2007,Richard2010}.
In particular, merging clusters present some of the strongest evidence for the existence of DM. The now iconic Bullet Cluster \citep{clowe2004,Bradac2006} is a well-studied example of a collision between two galaxy clusters, showing evidence for a clear separation between cluster members and the intra-cluster gas. Thanks to the phenomenon of gravitational lensing, which refers to the bending of the light near massive objects, it has been found that the majority of the mass was contained in the cluster DM halo, instead of the gas or the cluster members, as suggested by alternative theories of gravity \citep{Clowe2006}. Analogous studies have strengthened this result by showing that several other merging clusters present similar offsets \citep{Bradac2008,merten2011,harvey2015,Jauzac2016}.

Gravitational lensing has proven to be very efficient in constraining the mass distribution of galaxy clusters as it is independent of their dynamical states (e.g. \citealt{Kneib2012,hoekstra2013,bartelmann&maturi2017} for detailed reviews).
In the high mass density regions of the cluster core, the strong lensing regime, characterised by large distortions and multiple images of background sources, allows for a high-resolution mapping of the mass distribution. Different modelling techniques have been developed to take benefit of that effect and reconstruct the mass of cluster cores.

Cluster modelling methods based on strong lensing can be divided into two different classes. Methods belonging to the first class are based on physically motivated analytical mass models, they are usually referred to as \textit{parametric} methods. The most commonly used software for these \textit{parametric} methods are \textit{GLAFIC}\footnote{\url{https://www.slac.stanford.edu/~oguri/GLAFIC/}} \citep{oguri2010}, \textit{Lenstool}\footnote{\url{https://git-cral.univ-lyon1.fr/lenstool/lenstool}} \citep{jullo2007} and \textit{LTM} \citep{Zitrin2009}. They reconstruct the overall mass distribution by breaking it down into two type of components:
\begin{itemize}
    \item Cluster-scale components representing the mass contained in the cluster DM haloes and the gas in the intra-cluster medium.
    \item Galaxy-scale components representing the mass of each cluster galaxy member (the stellar mass as well as possibly attached DM halo)
\end{itemize}
Each component is modelled with a finite number of analytical density profiles such as 
the Singular Isothermal Sphere potentials (SIS, \citet{binnet&tremaine1987}), 
the symmetric power-law surface density profiles \citep{Broadhurst2005,Zitrin2009}, the Navarro-Frenk-White potentials (NFW, \citet{Navarro1997}), the dual Pseudo Isothermal Ellipsoidal Mass Distribution potentials (dPIE, \citealt{eliasdotti2007}) or the Pseudo-Jaffe ellipsoidal profiles \citep{Keeton2001}.  

Algorithms belonging to the second class are referred to as \textit{free-Form} (or also \textit{non-Parametric}) methods. They usually decompose the cluster mass distribution in a mesh of basis functions. The coefficients and possibly the shape of these functions are optimised to best reproduce the strong lensing constraints. Examples of software using these methods include \textit{SWUNITED} \citep{Bradac2005,Bradac2009}, \textit{GRALE}\footnote{\url{https://research.edm.uhasselt.be/jori/page/Physics/Grale.html}}\citep{Liesenborgs2006,Liensenborgs2009},\textit{WSLAP} \citep{diego2005,diego2007}, \textit{SAWLens}\footnote{\url{https://julianmerten.net/codes.html}} \citep{merten2009,merten2011} and \textit{LENSPERFECT}\footnote{\url{https://www.stsci.edu/~dcoe/LensPerfect/}} \citep{Coe2008,Coe2010}.

Both classes have their advantages and drawbacks. Although \textit{parametric} methods give physical models by construction, the use of analytical profiles is a strong assumption on the shape of the mass distribution, and this technique may have reached its limits in modelling some of the most complex galaxy clusters \citep{mahler2018}. At the other end of the spectrum, \textit{free-form} methods are by construction much more flexible to fit the observed data, but they need a much higher number of constraints to match their large degrees of freedom. As a consequence, their high flexibility may lead to non-physical mass distributions, and because of their tendency of over-fitting their prediction and estimation of physical quantities outside of the constrained area will be poor \citep{Coe2008}.  
While \textit{parametric} methods were initially the most-suited due to the small number of constraints accessible through strong lensing \citep{diego2005,kneib1993}, the number of constraints has continuously increased with time with the improvements of observations that culminated in imaging with the Hubble Frontier Fields (HFF) survey \citep{lotz2017} and in spectroscopy with the Multi Unit Spectroscopic Explorer (MUSE) \citep{Bacon2010} instrument that has dramatically increased the number of spectroscopically-confirmed multiply-imaged systems \citep{lagattuta2019}. Indeed, while pre-HFF studies on the cluster Abell 2744 provided less than ten spectroscopically-confirmed multiple images \citep{richard2014,johnson2014,merten2011}, HFF images combined with MUSE follow-up bring that number to almost one hundred \citep{mahler2018}. 

As part of the HFF initiative, multiple teams were invited to model the six HFF clusters. This effort provided material for an unprecedented comparison of these different techniques \citep{remolina-gonzalez2018,chirivi2018,raney2019}.  \citet{Priewe2017} inspected the magnification bias in the core of two HFF clusters and found a high dispersion between the different models from the third version. Simulated clusters representative of the  HFFs were used by \citet{meneghetti2017} to probe the different methods in order to determine their accuracy, and more recently \citet{rainey2020} studied the dispersion of the results among the different techniques on the mass profiles and the magnification on all six clusters. Notably, they show that mass profiles are consistent among the different techniques with only a $1\sigma$ scatter and a difference often below $5$ per cent. 

With the level of complexity reached by the HFF cluster models, \textit{parametric} methods seem to be now dominated by systematics \citep{mahler2018} and further developments are required on these techniques. The current number of constraints available call for the rise of hybrid methods which will enhance the \textit{parametric} modelling with a \textit{free-form} component. The recent example of \textit{Hybrid}-Lenstool use a large scale \textit{free-form} component to model clusters in their outskirts fitting weak-lensing measurements \citep{jullo2014,niemiec2020}. However, this method has only a \textit{parametric} modelling in the cluster core where the strong lensing constraints are available. An other example is WSLAP+ \citep{sendra2014} which introduces a \textit{parametric} component in a method that was originally only \textit{free-form}, to model the galaxy-scale components.

In this paper, we introduce a \textit{free-form} perturbative approach to the \textit{Lenstool} \textit{parametric} method to improve the strong lensing modelling. Thus, this method is complementary to the aforementioned improvements. In order to validate our approach, we tested it on a realistic galaxy cluster simulation representative of the Hubble Frontier Field clusters \citep{meneghetti2017}. Section~\ref{sec:method} recalls the \textit{parametric} approach implemented in \textit{Lenstool} and details the new \textit{free-form} approach. Section~\ref{sec:simulated} introduces the simulated cluster and the methodology followed for its modelling. In Section~\ref{sec:results}, we detail the obtained results and compare the parametric-only approach to the new hybrid one. Section~\ref{sec:discussion} quantifies the improvements on the reconstruction with a metric based on the data of the simulated cluster, and propose a method to select the appropriate \textit{free-form} modelling with a Bayesian criterion. Our conclusions are summarised in Section~\ref{sec:summary}.

We adopt a flat $\Lambda$CDM cosmology with $\Omega_m=0.3$, $\Omega_{\Lambda}=0.7$ and $H_0=70$ km$\text{s}^{-1}\text{Mpc}^{-1}$ throughout this paper.

\section{Method}
\label{sec:method}
\subsection{Strong lensing regime}
We consider a cluster to be a single plane lens under the thin lens approximation. The fundamental equation is then the lens equation mapping the source plane to the image plane (see \citealt{shneider1992} for a more detailed description):
\begin{equation}
    \vec{\beta}=\vec{\theta}-\vec{\nabla}\psi(\vec{\theta})
    \label{eq:lens_eq}
\end{equation}
where $\vec{\beta}$ and $\vec{\theta}$ are the angular position of the source and of the image, respectively. They are related through the gradient of the lensing potential $\psi$ computed at the position of the image. $\vec{\nabla}\psi$ is also referred to as the reduced deflection angle. In the case of multiply-imaged systems, this equation is degenerate and have multiple solutions $\vec{\theta}$ for a unique source position. This is happening in the strong lensing regime in the cluster core. The identification of such phenomenon allows an accurate mapping of $\psi$ in the cluster core, which is where we focus our reconstruction on.

The lensing potential $\psi$ is linked to the projected mass on the plane of the lens through the normalised surface mass density $\kappa$ (or convergence) defined as:
\begin{equation}
   \kappa=\frac{\Sigma}{\Sigma_{\rm crit}}= \frac{1}{2}\vec{\nabla}^2\psi
\label{eq:psi_poiss}
\end{equation}
where $\Sigma$ is the surface mass density of the lens and $\Sigma_{\rm crit}$ is the lensing critical surface density of the Universe. $\Sigma_{\rm crit}$ is defined as:
\begin{equation}
   \Sigma_{\rm crit}=\frac{D_s D_l}{D_{\rm ls}}\frac{c^2}{4 \pi G}
   \label{eq:sig_crit}
\end{equation}
where $D_l$, $D_s$ and $D_{\rm ls}$ represent the distances between the observer and the lens, between the observer and the source, and between the lens and the source, respectively. The magnification $\mu$ induced on background objects is equal to the determinant of the Jacobian matrix $M$
between $\vec{\beta}$ and $\vec{\theta}$:
\begin{equation}
    M_{i,j}=\frac{\partial\theta_i}{\partial\beta_j}
\end{equation}
Then we have:
\begin{equation}
    \mu={\rm det}(M)
\end{equation}

\subsection{Parametric modelling}
\label{sec:lenstool_meth}
The \textit{parametric} method implemented in \textit{Lenstool} is described in detail in \citet{jullo2007}. Briefly, the lensing potential is decomposed into cluster-scale and galaxy-scale components. Each cluster member is associated with a galaxy-scale component in the form of an analytical profile such as a dPIE \citep{eliasdotti2007}. Geometrical parameters such as the centre of cluster members on sky ($x,y$), projected ellipticity $e$ and position angle $\theta$ are measured from their light distribution. Thus, for a dPIE profile, the remaining parameters are the core radius $r_{\rm core}$, the cut-off radius $r_{\rm cut}$ and the central velocity dispersion $\sigma_0$.

Since the number of strong lensing constraints is small compared to the number of cluster members, each galaxy-scale component cannot be constrained individually. A global mass-to-light relation is used to link the mass with the luminosity of cluster members based on the \citet{Faber&Jackson1976} relation. We use this relation and a reference galaxy with a luminosity $L^*$ to obtain the following scaling relations:
\begin{equation}
\left\{
\begin{matrix}
r_{\rm cut}=r_{\rm cut}^*\left(\frac{L}{L^*}\right)^{1/2}, \\
r_{\rm core}=r_{\rm core}^*\left(\frac{L}{L^*}\right)^{1/2}, \\ 
\sigma_0=\sigma_0^*\left(\frac{L}{L^*}\right)^{1/4}
\end{matrix}\right.
\label{eq:scaling_rel}
\end{equation}
$L$ is the luminosity of the cluster member and $\sigma_0^*$, $r_{\rm core}^*$ and  $r_{\rm cut}^*$ are the dPIE parameters of the reference galaxy. $L^*$ is a characteristic luminosity where the galaxy luminosity function cuts off, and is chosen following the elliptical galaxy luminosity function at the cluster redshift \citep{schechter76}. 

Following observational modelling, the precise value of $r_{\rm core}^*$ has only a little effect on strong lensing predictions. We keep it fixed at $0.15$~kpc \citep{Limousin2007}. This only leaves to optimise $\sigma_0^*$ and $r_{\rm cut}^*$ to define all cluster member haloes. In the case of the Brightest Galaxy Cluster (BCG) or other particular galaxies, these relations could be relaxed. $r_{\rm cut}$ and $\sigma_0$ associated with these galaxies are optimised independently of $\sigma_0^*$ and $r_{\rm cut}^*$.

The cluster-scale components are also modelled with dPIE profiles, as they are mainly representing the dark matter we do not take any assumption other than $r_{\rm cut}=1$~Mpc as it is unconstrained by multiple images. The number of large-scale dPIE potentials used depends on the cluster geometry, but only a few are necessary, typically less than five.

\subsection{Perturbative modelling}
\label{sec:pert_modelling}
\subsubsection{Interest of B-spline surfaces}
B-splines are extensively discussed in the computer-aided design literature \citep{deBoor1978,NURBS-book,book-spline1,book-spline2}. They have the simplicity of polynomials functions and are only non-zero on finite supports. This avoids the Runge phenomenon \citep{Runge1901} of hyper-oscillation at high order and the non-locality of polynomial families.

A B-spline surface is defined by a set of basis functions separately or as a unique object defined below by its knots and basis function coefficients. Knots refer to points on the surface where the piece-wise polynomial are connected. With this last formalism, it is possible to compute them efficiently with the \citep{deBoor1978} algorithm detailed in Appendix~\ref{sec:comp_Bspline}.

One reason B-splines have a superior computational efficiency over most of multi-scale grids of potentials (e.g. \citet{jullo2009}) is that they have a finite support, so they are not computed for every evaluation of $\psi$ or its derivative. A second reason is the algorithmic complexity of the De Boor algorithm for evaluating a point, which only depends on its polynomial degree and not on the number of basis functions. These properties are conserved for all derivatives as they are also B-spline surfaces.

\subsubsection{Constructing the B-spline surface}
A B-spline surface is defined as a tensor product of one-dimensional B-spline curves. As these curves are piece-wise polynomials there exists a set of points called knots which indicate the connection between each polynomial piece. Thus, we have two sets of knots $(t_{x,i})_{i\in[\![1,N]\!]}$ and $(t_{y,i})_{i\in[\![1,N]\!]}$ where $N$ is the number of knots in each direction, and $x$ and $y$ refer to orthogonal axes in the lens plane.

The B-spline surface $\Delta\psi(x,y)$ is expressed as a function of one dimensional B-spline basis functions $B_{j,p,t}$ where $j$ is the knot index, $t$ is the knot vector and $p$ is the polynomial degree. We have two knot vectors $t_x$ and $t_y$, one for each axis which leads to the following formula:
\begin{equation}
    \Delta\psi(x,y)=\frac{D_{ls}}{D_{s}}\sum^m_{j,l=1} C_{j,l} B_{j,p,t_x}(x) B_{l,p,t_y}(y)
    \label{eq:pert_surf}
\end{equation}
where $C_{j,l}$ are the coefficients of each B-spline basis and $m$ is the number of B-spline functions per axis. We apply the factor $D_{ls}/D_{s}$ to keep the scaling specific to the lensing potential. Thus, we have $m^2$ basis functions and $m$ is related to $N$ and $p$ by:
\begin{equation}
    N=m+p+1
\end{equation}
and each B-spline basis $B_{j,p,t_x}(x)$ is only non-zero for $t_{x,j}\la x < t_{x,j+p+1}$ and can be expressed recursively by the Cox-De Boor recursion formula \citep{deBoor1978}. The same relation holds for the basis defined on the second axis. More information on the current implementation of the B-spline calculation can be found in Appendix~\ref{sec:comp_Bspline}.

\begin{figure}
    \centering
    \includegraphics[width=\linewidth]{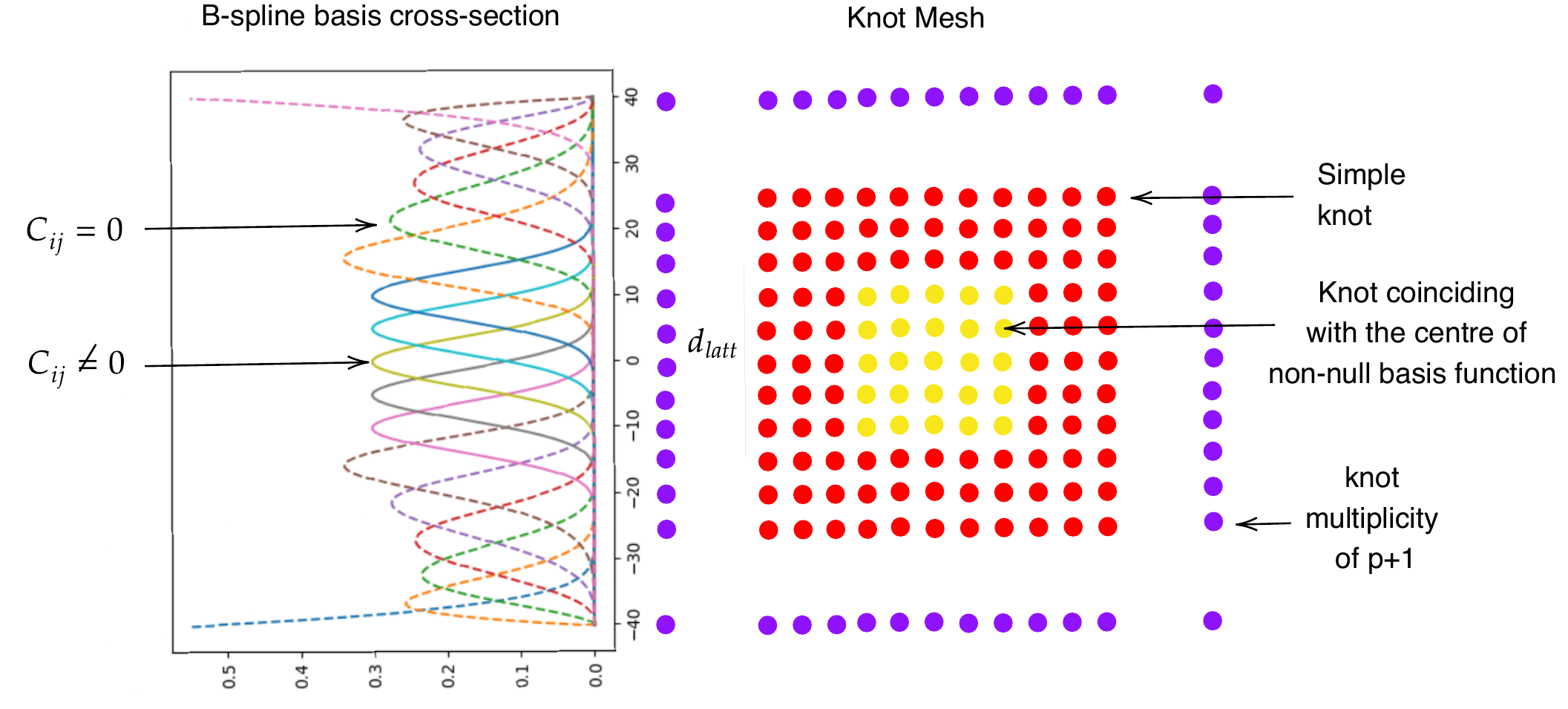}
    \caption{Overview of the B-spline surface parametrisation used in our perturbative modelling. The diagram shows a cross-section of B-splines basis functions with the corresponding knot mesh, illustrated for a polynomial degree $p=5$. Purple, red and green dots represents knots with a multiplicity of $p+1$ (i.e. $p+1$ knots sharing the same position), simple knots (i.e. multiplicity of $1$) and simple knots that are coincident with the centre of non-null basis functions, respectively.}
    \label{fig:meshknot}
\end{figure}

Moreover, to define the knots on each axis we use the "averaging" techniques in an analogous manner to the formalism of interpolating B-splines \citep{NURBS-book,deBoor1978}. For both axes, we define the points $(C^x_{j,l},C^y_{j,l})_{j,l\in [\![1,m]\!]^2}$ placed on a regular mesh. Considering only the x-axis this leads to the following expression for the knots that can be generalised to the y-axis:
\begin{equation}
    t_{x,i+p}=\frac{1}{p}\sum^{i+p-1}_{j=i} \left(\frac{1}{m}\sum^m_{l=1}C^x_{j,l}\right) \text{ for $i=1,...,m-p$}
\end{equation}
The remaining knots are defined as:
\begin{align}
    &t_{x,i}= \text{min}\left(\frac{1}{m}\sum^m_{l=1}C^x_{j,l}\right)\text{ for $i=0,..,p$}\\
    &t_{x,i}= \text{max}\left(\frac{1}{m}\sum^m_{l=1}C^x_{j,l}\right) \text{ for $i=m+1,..,m+p+1$}
\end{align}
These two paddings ensure that the B-spline surface $\Delta\psi(x,y)$ will be equal to the $C_{j,l}$ values at the surface limits. The same formula applies for $t_{y,i}$ with $C^y_{l,j}$ instead of $C^x_{j,l}$. Even if $(C^x_{j,l},C^y_{j,l})_{j,l\in [\![1,m]\!]^2}$ are regularly spaced, the knot mesh is only regular at the centre of the surface. Fig.~\ref{fig:meshknot} shows the resulting knot mesh for $p=5$ and a cross-section of the B-spline basis functions. If $p$ is an odd integer and the knots are regularly spaced, the centres of the B-spline basis are coincident with knots as their supports are squares with a side size of $p+1$ times the distance between two consecutive knots. 

\subsubsection{Assumptions on the surface}
There are several options to add a B-spline surface on models. The first and easier is to add it to the lensing potential $\psi$. All physical quantities related to the lensing can be computed from $\psi$ or its derivatives, in the case of a B-spline surface, all these functions are analytical. The only condition is to ensure that the resulting mass distribution is plausible, but for our perturbative approach, we only have to ensure that the added modifications are small compared to the \textit{parametric} component.

We decided to add the B-spline surface perturbation $\Delta\psi$ onto the lensing potential $\psi$ from the parametric-only modelling. All derived quantities (deflection, magnification) are in that case analytical and fast to compute. This approach differs from perturbing the mass distribution \citep{Broadhurst2005} or the deflection field \citep{Coe2008}.

We impose the following conditions to ensure the continuity of $\Delta\psi$ and its derived quantities, as well as to connect the B-spline surface to the parametric $\psi$:
\begin{itemize}
    \item A polynomial degree $p\geq 4$.
    \item $\Delta\psi(x,y)=0$ and $\Delta\psi^{(j)}(x,y)=0$ for $j=1,..,p$ and $x\in\{\text{min}(C^x_{j,l});\text{max}(C^x_{j,l})\}$ or $y\in\{\text{min}(C^y_{j,l});\text{max}(C^y_{j,l})\}$
\end{itemize}
The first condition ensures that inside its area of definition, the first three derivatives of the potential $\Delta\psi$ (the deflection angle, the convergence and the shear, and the flexion components) are continuously differentiable, its fourth derivative being only piece-wise continuous. The second one allows the surface to be extrapolated to a value of zero outside its limit with a connection that has the same continuity as the surface on its knots. To fulfil this condition we impose the following restriction on the B-spline coefficients $C_{j,l}$:
\begin{equation}
C_{j,l}=0 \text{ for }(j,l)\in\{1;..;p+1;m-p;..;m\}^2
\end{equation}
As $\Delta\psi$ and its derivatives are B-splines surfaces, the padding at the beginning and at the end of the knot sequence ensures that these surfaces will be equal to their associated coefficients at their edges. Hence, it is straightforward to see that this condition implies $\Delta\psi(x,y)=0$ at the mentioned points. Basis coefficient of the $\Delta\psi^{(j)}$ are linear combinations of the $C_{j,l}$. The coefficient of the basis functions of $\Delta\psi^{(j)}$ on the border of the patch are zero due to the condition imposed on the $C_{j,l}$. Thus, $\Delta\psi^{(j)}$ are also null at the patch limit. As shown in Fig.~\ref{fig:meshknot}, we define $d_{\rm latt}$ which represents $n+1$ times the regular space between two $(C^x_{j,l},C^y_{j,l})$ with $n$ representing the number of non-zero B-spline basis functions on a line of the knot mesh. In this example and given the continuity conditions described in this section, in order to construct a B-spline surface $\Delta\psi(x,y)$ of degree $p=5$ and $n=5$, we have $m=17$ and $N=23$ such that there is a total of $m\times m=289$ basis functions of which the amplitude of only $n\times n=25$ basis functions is not set to zero and is optimised. The total number of knots is $N\times N=529$, of which $(m-p-1)\times(m-p-1) = 121$ are simple knots. We can also relate $m$, $n$ and $p$ thanks to the continuity conditions by:
\begin{equation}
    m=n+2(p+1)
\end{equation}

Because $\Delta\psi(x,y)$ is extrapolated to zero outside the perturbation limits, the total mass added by the perturbation is always zero. Hence, the perturbation is only deforming the mass distribution, but the total mass is equivalent to the one contained in the dPIEs only. As the mass profiles are well-constrained by strong lensing in the cluster cores, this is not an issue \citep{rainey2020}.

\subsection{Modelling process}
The free parameters are estimated through a Monte Carlo Markov Chain (MCMC) process by the Bayesian engine \textit{bayeSys} \citep{Skilling2004} implemented in \textit{Lenstool} by maximising a Likelihood function.
\subsubsection{Likelihood definition}
\label{sec:likelihood}
The most simple strong lensing likelihood implemented in \textit{Lenstool} assumes multiple images to be point-like objects. This likelihood is defined in the lens plane as follows:
\begin{equation}
 \mathfrak{L}_{\rm SL}=\prod_{i=1}^N \frac{1}{\prod_{j}\sigma_{i,j}\sqrt{2\pi}}\exp{\left(-\chi_i^2/2\right)}
 \label{eq:likely-s}
\end{equation}
With $\chi_i^2$ defined for the $i^{\rm th}$ system as:
\begin{equation}
\chi_i^2=\sum_{j=1}^{n_i}\frac{(\vec{\theta}^{\rm obs}_j-\vec{\hat{\theta}}_j)^2}{\sigma_{i,j}^2}
\end{equation}
where $\vec{\theta}^{\rm obs}_j$ is the observed position of the image, $\vec{\hat{\theta}}_j$ is the model predicted position and $\sigma_{i,j}$ is the error on the $j^{\rm th}$ image of the $i^{\rm th}$ system. $\sigma_{i,j}$ incorporates the measurement error on the image centroid but also all uncertainties from perturbations on the line of sight. Values for $\sigma_{i,j}$ are frequently chosen from $0.2$~arcsec and up to $1$~arcsec. Here, we choose it equal to $0.2$~arcsec and constant for all multiple images \citep{host2012,jullo2010,DAloisio2011}. It is not an issue to make such choice as all multiple images used in this paper are calculated with the true mass distribution without added uncertainties. Model predicted positions $\vec{\theta}^{\rm obs}_j$ are obtained by computing the multiple images from the barycentre of sources of each observed position. To make this barycentre more reliable each source position is weighted by the corresponding magnification.

\subsubsection{MCMC sampling}
\label{sec:MCMC-sampl}
The prior distributions of the free parameters can be either uniform or normal. In the case of the perturbative potential, the free parameters are the centre coordinates and the position angle $\phi$ of the knot mesh, the lattice size $d_{\rm latt}$ and the coefficient $C_{j,l}$ describing the amplitude of the perturbation. 

The prior distribution for the position angle $\phi$ is uniform between $0^\circ$ and $180^\circ$. Uniform priors on the mesh centre and on the lattice size are chosen so that the perturbation could affect the predicted position of all multiple images. The centre position of the mesh is allowed to vary within $4$~arcsec from the centre of a circle enclosing all constraints. The radius of this circle is then used to define a lower bound for the lattice size $d_{\rm latt}$, while the higher bound is typically chosen to be $20$ per cent larger. Wider priors on $d_{\rm latt}$ tend to cause issues with the MCMC engine which struggles to converge to high likelihood areas.

In order to avoid the perturbation to dominate the parametric potential and produce a non-physical mass distribution, we chose to use a zero-mean Gaussian prior on the $C_{j,l}$ coefficients. The standard deviation of the Gaussian prior is equal for all the coefficients, and its value depends on the $d_{\rm latt}$ range because both values of coefficients and $d_{\rm latt}$ are the ones that define the possible mass added or subtracted by the perturbation. Since the B-spline coefficients $\frac{\partial^2\Delta \psi}{\partial x^2}$ or $\frac{\partial^2\Delta \psi}{\partial y^2}$ are linear combinations of the $C_{j,l}$, this allows us to estimate the maximum values for these derivatives assuming $D_{\rm ls}/D_{s}=1$ by:
\begin{equation}
    \text{max$\left(\frac{\partial^2\Delta\psi}{\partial x^2},\frac{\partial^2\Delta \psi}{\partial y^2}\right)$}\la \frac{4\text{max$\left(C_{j,l}\right)$}}{(d_{\rm latt}/n)^2}
\end{equation}
We choose to use max$\left(C_{j,l}\right)\sim2.7\sigma_{\rm prior}$ with $\sigma_{\rm prior}$ the standard deviation of the Gaussian prior probability. This assures us that $99.3$ per cent of the parameter values sampled from this prior will be below this limit. Hence, perturbed models with a different number of B-splines basis will have a comparable effect on the cluster mass distribution. Finally, we note that the use of a zero-mean Gaussian prior effectively acts as a regularisation of the amplitude of the perturbation.

Even with this choice of priors, it was difficult for \textit{bayeSys} to explore correctly the full parameter space and to converge towards high-likelihood regions. To overcome this convergence issue, we followed a 2-step approach. First, we optimise a parametric model using uniform priors on parameters and do not include a perturbation. This gives us a fiducial model which serves as a baseline for the comparisons discussed in Section~\ref{sec:results}. Then, we use this run to define new Gaussian priors for these parameters. These priors are centred on the value of the best models, and the standard deviation is chosen to be three times the standard deviation of the distribution given by the MCMC, while keeping relevant parameters in a physical range. This speeds up the next run as the model starts directly in regions with a high likelihood. If the chains converge towards a prior tail, it will be modified to better explore specific parameters. This process can be iterative to avoid a bias induced by the first run on the priors.

In summary, this 2-step approach illustrates the founding principle of our modelling method. First, an approximate model is found using a well-established \textit{parametric} method. Then, this model is further refined with the help of a \textit{free-form} perturbation.

\section{Tests on a simulated cluster}
\label{sec:simulated}
\subsection{Description of the simulation}
The HFF modelling challenge was launched in 2013 with the goal of comparing different modelling techniques (parametric, free-form, and hybrid) on HFF-like clusters. For this purpose, two different galaxy cluster simulations were produced and made publicly available\footnote{\url{http://pico.oabo.inaf.it/~massimo/Public/FF/hera.html}\label{foot:model_meneghetti}} \citep{meneghetti2017}. 

Both clusters were simulated using different techniques. The resulting bi-modal mass distributions are typical of complex massive galaxy clusters in a flat $\Lambda$CDM cosmology. We refer the reader to \citet{meneghetti2017} for a detailed description of these clusters. Here, we briefly summarise the properties that are relevant to our work. The first one, \textit{Ares}, is a semi-analytical cluster \citep{Giocoli2012} where \textit{parametric} methods performed well in its reconstruction. This is partially due to the use of analytical profiles in the simulated cluster that are similar to the ones used to reconstruct the mass distribution. The second one, \textit{Hera}, is taken from an N-body simulation of cluster-sized DM haloes with only collisionless DM particles \citep{plantelles2014}. The accurate modelling of this simulated cluster was found to be much more difficult for all the methods tested in the challenge.
Especially, dPIE profiles are not well-suited to represent cluster-scale components as large-scale simulations show an agreement with NFW profiles \citep{deBlock2009}. In the case of galaxy-scale components, they are in agreement with mass density obtained through galaxy-galaxy lensing measurements \citep{bolton2008,shu2016}. Indeed, with a negligible $r_{core}$ as in our case, they reproduce a quasi-isotherm profile. However, there is still a discrepancy with profiles obtained from simulated data, as they present a less concentrated distribution a their centre \citep{Meneghetti2020}.
\citet{acebron2017} showed that even with different combinations of analytical profiles, the \textit{parametric} method of \textit{Lenstool} is unable to reduce the error on the reconstruction of multiple images positions.

We focus our tests of the perturbative approach on the \textit{Hera} cluster, as it presents an unfavourable case for the traditional method and offers great potential for improvements.

\begin{figure*}
    \begin{minipage}{0.49\linewidth}
    \centering
    \includegraphics[width=\linewidth]{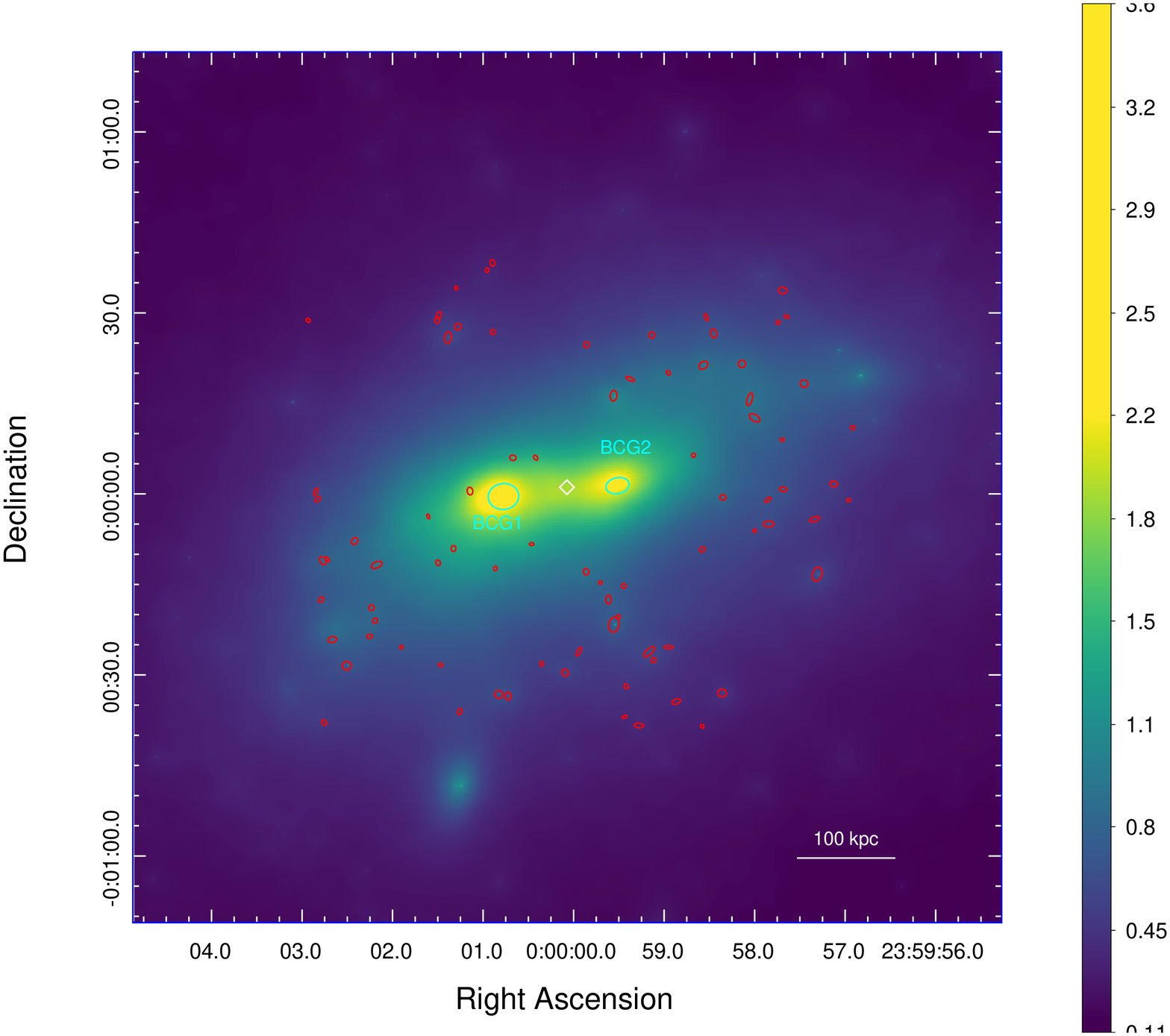}
    \end{minipage}
    \begin{minipage}{0.49\linewidth}
    \centering
    \includegraphics[width=0.9\linewidth]{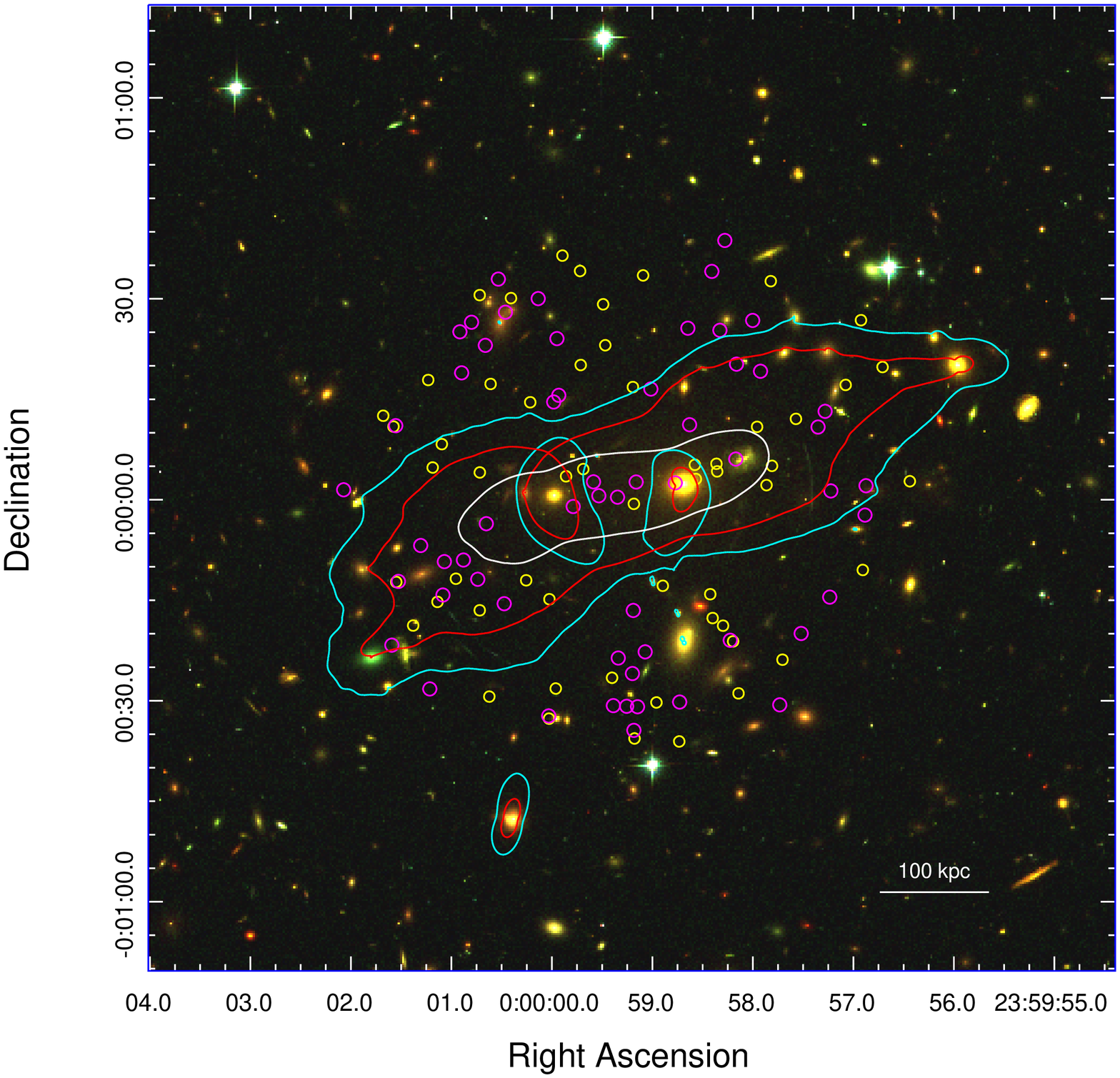}
    \end{minipage}
    \caption{\textit{Left panel}: true convergence distribution of the \textit{Hera} simulated cluster core, the red ellipses represent the cluster members modelled according to the scaling relations, the cyan ones the two BCGs. The white diamond marks the central location selected to compute the mass profiles. \textit{Right panel}: colour HST-like image of the simulated data with the first and second sets of multiple images represented by yellow and magenta circles, respectively. Critical lines are shown for $z=1$(white), $2$(red) and $7$(cyan).}
    \label{fig:kappa_sim}
\end{figure*}
\begin{table}
	\begin{tabular}{lcccr}
	\hline
	Set & $N_{\rm sys}$ & $N_{\rm im}$ & $z_{\rm min}$ & $z_{\rm max}$\\
		\hline
		$1$ & $17$ & $55$ & $1.23$ & $6.58$\\
		$2$ & $18$ & $57$ & $1.20$ & $6.41$\\
		\hline
	\end{tabular}
		\centering
	\caption{Summary of the strong lensing constraints produced for our simulations. For each set, we provide the number of multiply-imaged systems $N_{\rm sys}$, the total number of multiple images $N_{\rm im}$, minimum ($z_{\rm min}$) and maximum ($z_{\rm max}$) source redshifts.}
	\label{tab:multiple-im}
\end{table}

Simulated observations and lensing observables were provided for the HFF challenge, but we decided to create our own sets of lensing data with the publicly available deflection field\textsuperscript{\ref{foot:model_meneghetti}}. This allows us to have more flexibility in the number and variety of realisations of multiple images. It also ensures that the cosmology is consistent through all the following analysis. Therefore, we created two different sets of multiple images by solving the lens equation (Eq.~\ref{eq:lens_eq}) for random source positions. Each set has multiple images that are distributed homogeneously in the strong lensing region. Both sets have similar numbers of systems and multiple images which are described in Table \ref{tab:multiple-im}. The redshift of the sources follows a uniform probability distribution in the range $1<z<7$. The positions of the different constraints are displayed in Fig.~\ref{fig:kappa_sim} where the background image is an HST-like observations of the cluster provided for the HFF challenge.

\subsection{Description of models}

The same methodology was applied to both sets of multiple images. Cluster members which follow the scaling relations are presented in Fig.~\ref{fig:kappa_sim}; they are chosen following the model produced by the Cluster As TelescopeS (CATS) team for the HFF challenge  \citep{meneghetti2017}. The two galaxies that dominate the cluster do not follow the scaling relations; they are labelled as BCG-1 and BCG-2 in Fig.~\ref{fig:kappa_sim}. All of these galaxies have $mag_{\rm AB}<24.0$ in the F$814W$ \textit{HST} filter. As the cluster is bi-modal, we add two dPIE profiles to reproduce the smooth component with a prior on their position centred on each BCG.

For the perturbative modelling, we tried to vary the number of B-spline basis on a uniform grid for all sets of constraints. We began with a minimum of $9$ B-splines in a mesh of $3$ per $3$. We increased the size of the mesh progressively to a maximum of $12$ per $12$; this makes a maximum of $144$ B-splines basis. For more convenience, we will label the perturbed models with their number of non-zeros basis functions per mesh line $n$ (e.g. Fig.~\ref{fig:meshknot}).

The centre of the B-splines mesh is allowed to vary uniformly in a square centred on $\Delta \alpha= 13.395$~arcsec, $\Delta \delta= 4.380$~arcsec ($\Delta \alpha$ and $\Delta \delta$ are computed relatively to the BCG-1 positions used as reference) with a side size of $8$~arcsec. $d_{\rm latt}$ is chosen with a uniform prior between $100$~arcsec and $120$~arcsec according to the description given in Section~\ref{sec:MCMC-sampl}. We choose the standard deviation of the $C_{j,l}$ such that the maximum amount of the convergence is $\kappa=1.75$ (assuming $D_{\rm ls}/D_{\rm s}=1$), this keeps the perturbed surface in a reasonable range compared to the parametric-only part.

\section{Results}
\label{sec:results}
We now compare the simulated data of the \textit{Hera} cluster to the fiducial and the perturbed models. We present in Section~\ref{sec:repro-im} results on the mapping of the deflection angle fields through the reproduction of the multiply-imaged systems. Differences in the 2D mass distributions and mass density radial profiles are shown in Section~\ref{sec:mass_dist} and Section~\ref{sec:profiles}, respectively. We also investigate the effective Einstein radii in Section~\ref{sec:Einstein_radii} and the distribution of the substructure masses in Section~\ref{sec:mass_sub}. Each comparison is made with models optimised on the same set of constraints.

\begin{figure*}
    \begin{minipage}{0.49\linewidth}
    \centering
    \includegraphics[width=\linewidth]{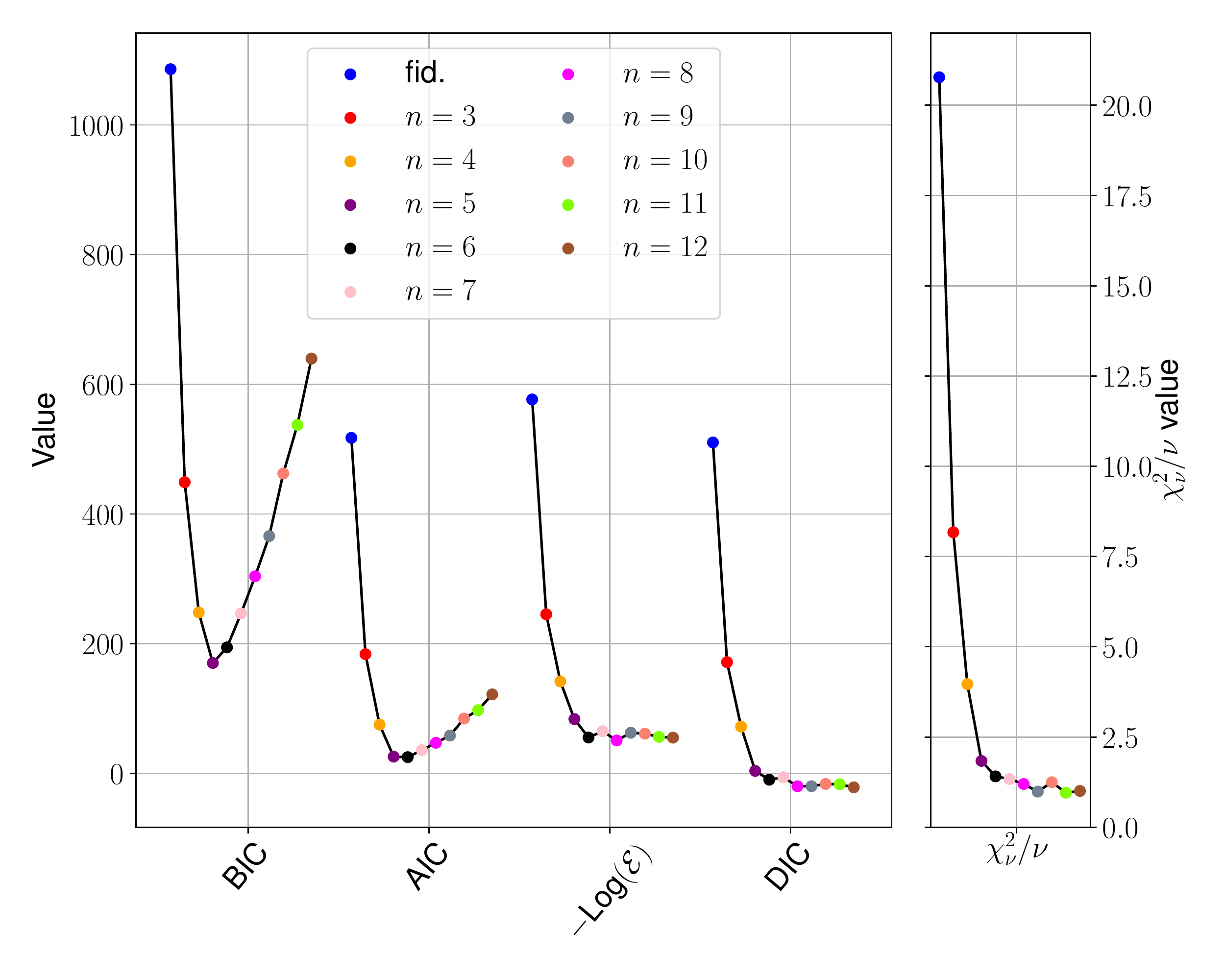}
    \end{minipage}
    \begin{minipage}{0.49\linewidth}
    \centering
    \includegraphics[width=\linewidth]{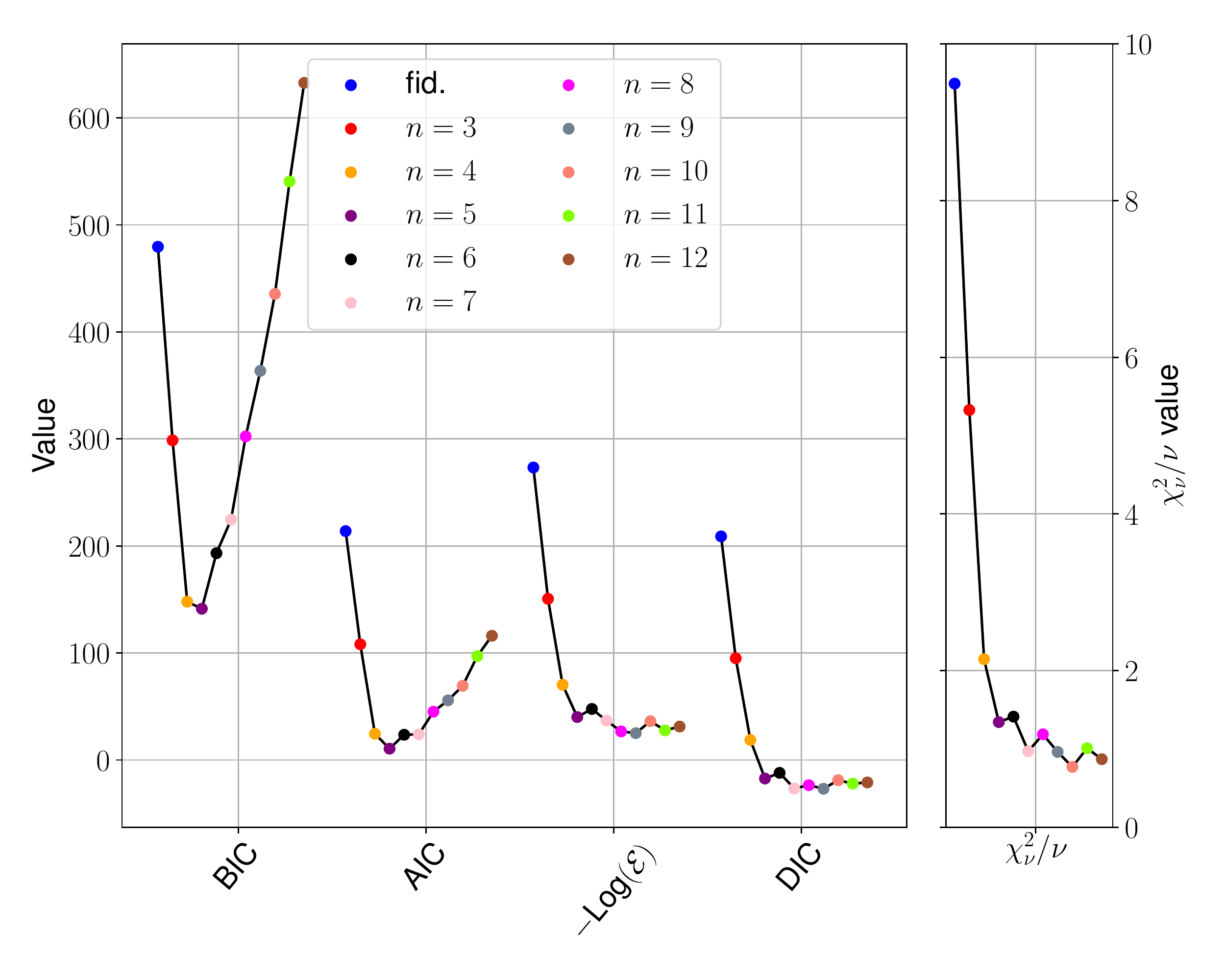}
    \end{minipage}
    \caption{\textit{Left panel}: Summary of values obtained for the $-$Log$(\mathcal{E})$, BIC, AIC, DIC and $\chi^2/\nu$ related to each model optimised on the first set of multiple images. \textit{Right panel}: Same plot but for models fitted on the second set of constraints.}
    \label{fig:bayes_indicator}
\end{figure*}

\subsection{Bayesian criteria}
\label{sec:criteria}
Our tests provide us with multiple models reproducing correctly the same constraints, but with different number of parameters. In order to distinguish between these models, we make use of different criteria based on information theory or Bayesian inference \citep{liddle2007}.

We use the likelihood function of the predicted multiple images positions and the number of parameters estimated by the model to compute the following criteria: Bayesian Information Criterion (BIC; \citealt{Schwarz1978}), Akaike Information Criterion (AIC; \citealt{Akaike1973}), Deviance Information Criterion (DIC; \citealt{Spiegelhalter2002}), and the negative log-Evidence $-$Log$(\mathcal{E})$ \citep{jullo2007}. Fig.~\ref{fig:bayes_indicator} compares the obtained values with the usual $\chi^2$ for best models, here divided by the number of multiple images $\nu$. We also investigated values given by the Widely Applicable Information Criterion \citep{Watanabe2010} and the Widely Applicable Bayesian Information Criterion \citep{Watanabe2013}. However, as they show similar results to the DIC and BIC, respectively, we do not show them on the graph for more clarity. We note that BIC, AIC and $\chi^2/\nu$ are computed for the best model only while the Evidence and the DIC both take into account the full posterior distribution of models. Most of these criteria except $\chi^2/\nu$ are meant to be compared between models. A model must show a strong difference in a given criterion to be judged better than another. In the case of the AIC, BIC and DIC, a difference of $10$ rules out a model compared to another and for the $-$Log$(\mathcal{E})$ we choose a difference equivalent to $5\sigma$ significance which is about $12.5$ \citep{trotta2008}. When two models are judged equivalent based on these conditions, we apply Occam's razor by selecting the model with the fewer parameters. To determine the best model according to $\chi^2/\nu$ we select models that reach a value of $1$ or less and apply Occam's razor.

Most notably, there are two different patterns among all criteria, AIC and BIC show a clear minimum with $n$, when the three others reach a plateau where all models are equivalent. Considering the rules introduced before, BIC and AIC favour models with $n=5$ for both set of constraints. We note that the AIC indicates that models with $n=5$ and $n=6$ are equivalent on the first set of constraints. The best models according to $-$Log$(\mathcal{E})$, DIC and $\chi^2/\nu$ have $n=6,12,9$ and $n=8,5,7$ for models optimised on the first and on the second set of constraints, respectively.

Regarding our rules, the DIC favours $n=12$ on the first set of constraints because it shows a significant difference on the DIC compared to $n=6$ (models with $6<n<12$ being equivalent to the latter). However, it is likely that these two models are also equivalent due to possible estimation error, which would imply that the model with $n=6$ is the best according to this criterion. Similarly, $-$Log$(\mathcal{E})$ shows the same pattern on the second set of constraints with $n=5$ and $n=8$. In both cases, the smaller $n$ is the one when these criteria reach their plateau. In conclusion, most criteria agree on the best value of $n$, whether they reach a plateau or the best models according to our rules. The $\chi^2/\nu$ tends to select models with a higher $n$ but it strongly depends on our choice on the observational error on the multiple images positions.

We will assess in the next section the accuracy of these criteria with metrics based on a comparison with the simulated data. The goal is to define a criterion to select $n$ in the case of a real cluster.

\begin{figure*}

    \includegraphics[width=\linewidth]{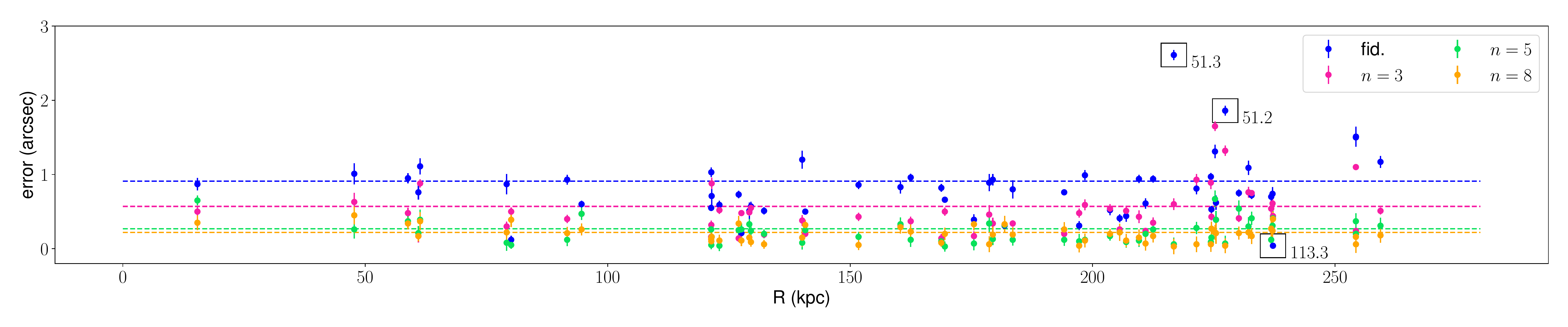}
    
    \includegraphics[width=\linewidth]{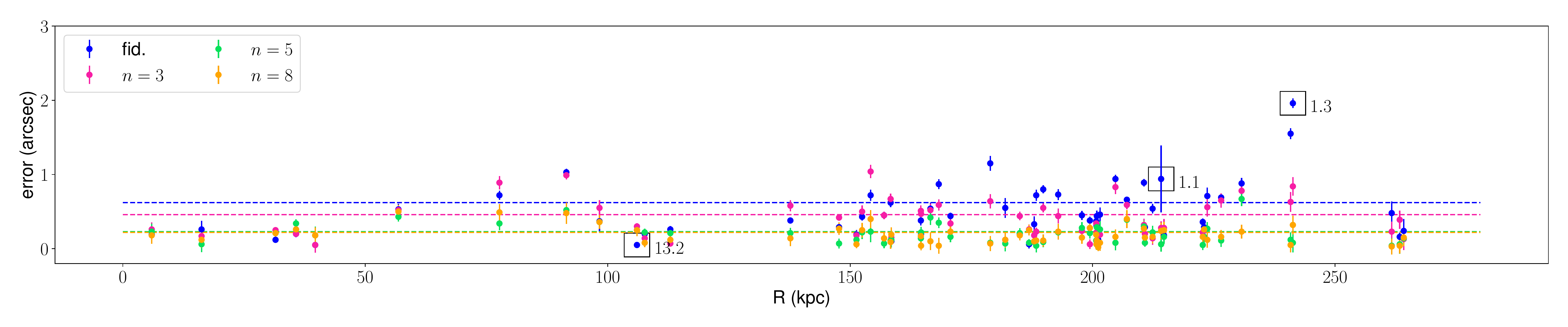}
    
    \caption{\textit{Top panel}: residual distances between the observed and the predicted multiple images positions. These distances are provided for the best-fitting models optimised on the first set of constraints. Errors show the statistical uncertainties from the MCMC runs. Multiple images are plotted in ascending order according to their projected distance to the cluster centre, converted to physical distance in the lens plane. The dashed lines represent the RMS achieved by a model among a specific sets. Each lines is associated to the model that is represented by the same colour on the scatter plot. Labelled multiple images are deviant points and are discussed in Sect~\ref{sec:repro-im}. \textit{Bottom panel}: Same plot but for the second set of multiple images.}
    \label{fig:error_im}
\end{figure*}

\subsection{Reproduction of multiple images positions}
\label{sec:repro-im}
The first improvement obtained by the addition of the B-spline surface is a better reproduction of the multiple images positions. This is shown in Fig.~\ref{fig:error_im} which presents the residual distances between the predicted and the observed multiple images. Error bars shown in these figures are the statistical uncertainties from the MCMC run. These figures show a clear improvement when we compare the fiducial models (which are parametric-only) to models with B-splines perturbations, and this enhancement is consistent among all multiply-imaged systems. We see that the root-mean-square (RMS) error on the multiple-image position decreases as we increase the number of B-spline functions ($n^2$) added to the models.

For the first set of constraints, we obtain an RMS of $0.20$~arcsec for $n=11$ on the best models which is an improvement by more than a factor of 4 when compared to the RMS of $0.91$~arcsec obtained for the fiducial model. Variants from $n=9$ to $n=12$ all reach the assumed observational errors of $0.2$~arcsec and show similar residual distance distributions without significant enhancement. The RMS of models from $n=3$ to $n=8$ are progressively decreasing as shown by the dashed lines in Fig.~\ref{fig:error_im}. For the first set of constraints, we obtained RMS values of $0.57$~arcsec, $0.27$~arcsec, and $0.22$~arcsec, for models with $n = 3$, $5$, and $8$, respectively. The corresponding RMS values obtained with the second set of constraints are $0.46$~arcsec, $0.23$~arcsec, and $0.22$~arcsec. The second set of constraints shows a similar trend, with perturbed models reaching the observational errors of $0.2$~arcsec at $n\geq7$ while the fiducial model had an RMS three times higher with $0.62$~arcsec.

If we look at multiple images more precisely, the fiducial models exhibit several multiple images that are poorly reproduced (e.g. $51.3$ and $51.2$ or $1.1$ and $1.2$ for the first and second set of images, respectively) but also images that are very well predicted within the assumed observational error (e.g. $113.3$ and $13.2$ for the first and second set of images, respectively). However, as $n$ is increasing, the spread of the distribution of these distances is reduced, making the reconstruction of multiple images more homogeneous. More quantitatively, the standard deviation among these distributions is reduced for both sets by a factor $\sim4$ between fiducial models and the perturbed ones with $n=7$ ($>0.35$~arcsec compared to $\sim0.10$~arcsec for both sets of constraints).

\begin{figure*}
    \begin{minipage}{0.49\linewidth}
    \centering
    \includegraphics[width=\linewidth]{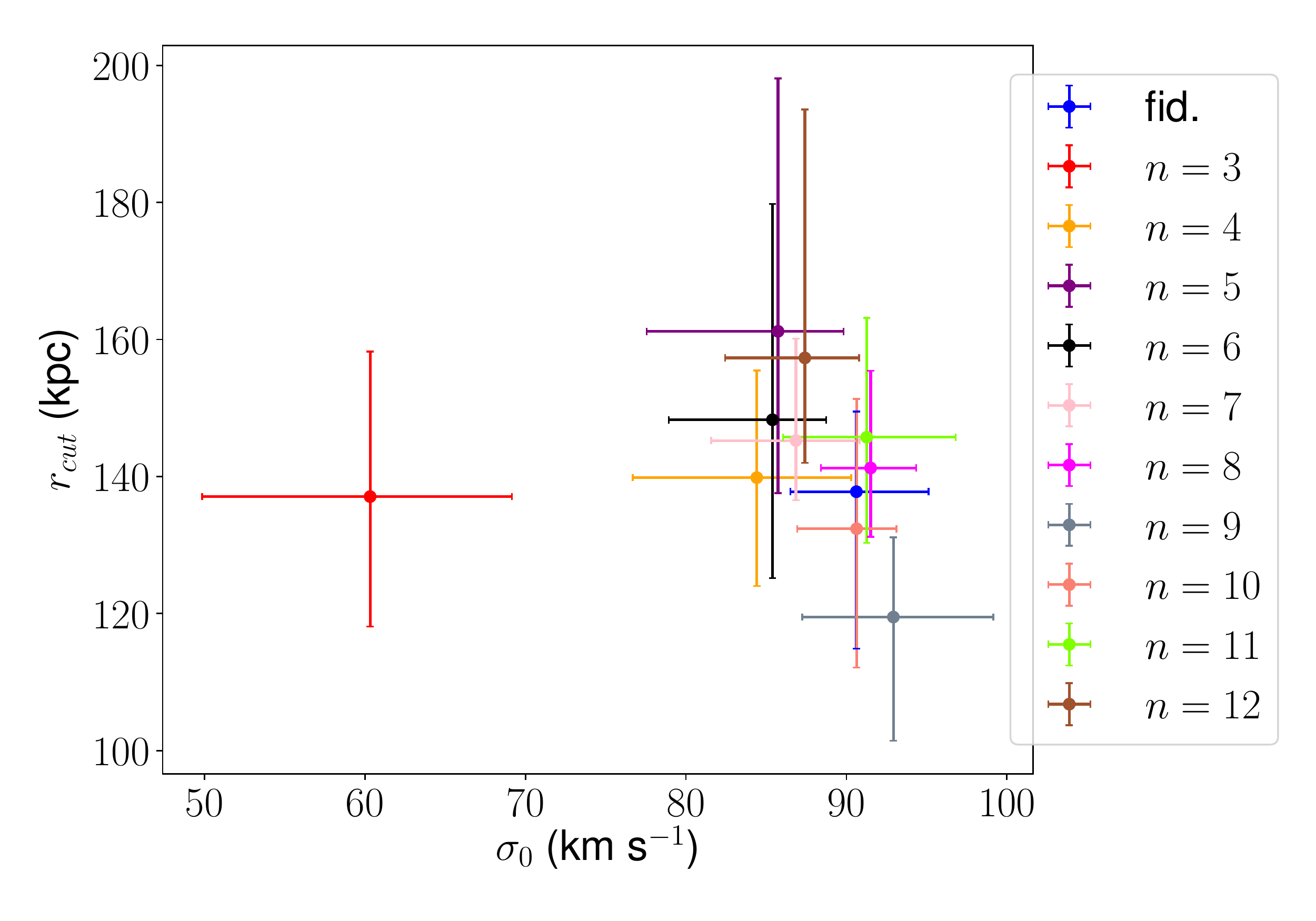}
    \end{minipage}
    \begin{minipage}{0.49\linewidth}
    \centering
    \includegraphics[width=\linewidth]{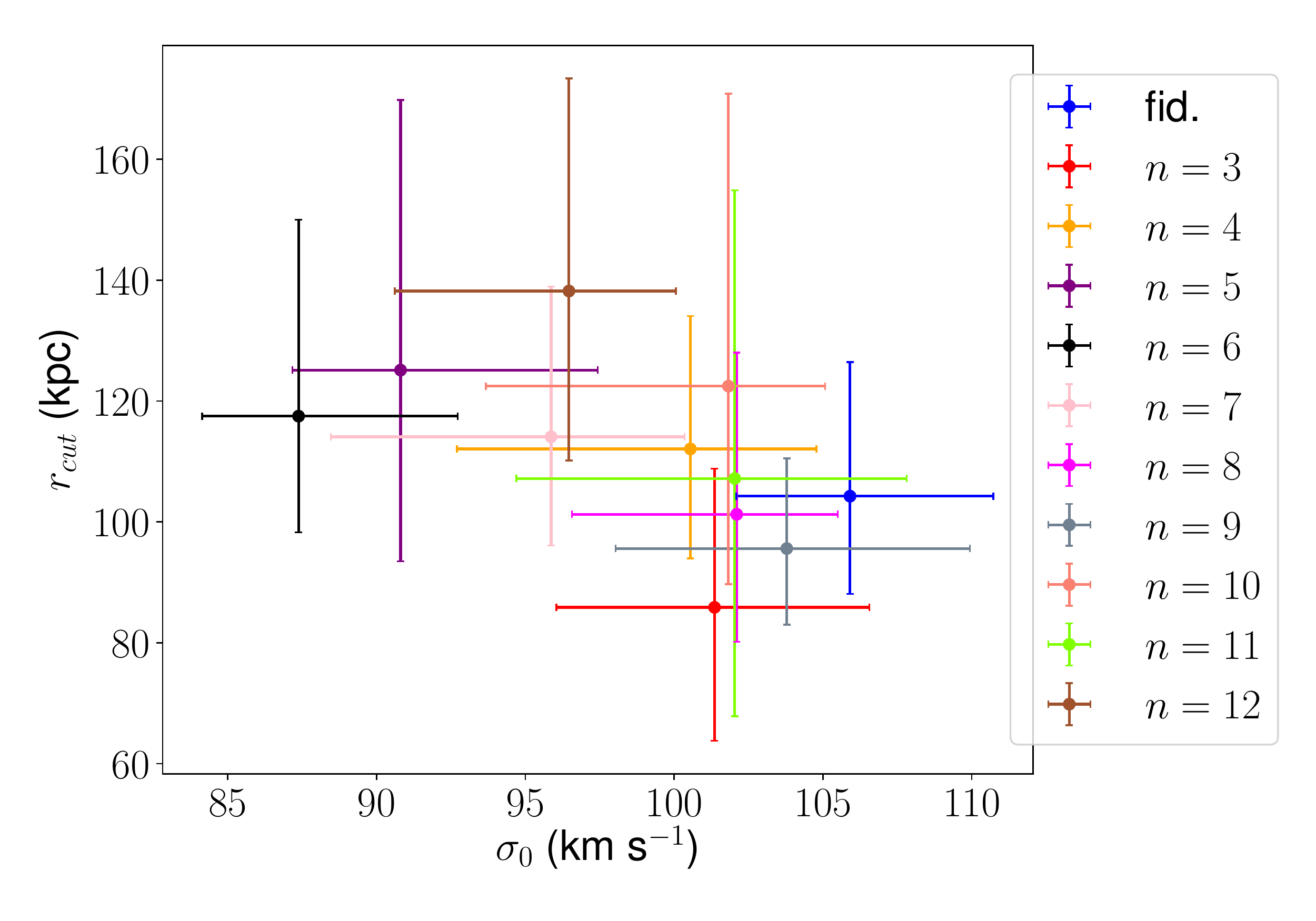}
    \end{minipage}
    
    \caption{\textit{Left panel}: Parameter space related to cluster member profiles. The cut radius $r_{\rm cut}^*$ is plotted in function of the central velocity dispersion $\sigma_0^*$ and the errors represent $68$ per cent of the distribution of these parameters from the MCMC output. Models considered are optimised on the first set of multiple images. \textit{Right panel}: same plot for the second set of constraints.}
    \label{fig:sigposVSrcut}
\end{figure*}

\subsection{Mass distribution}
\label{sec:mass_dist}
\subsubsection{Parametric components}
The parametric part of the mass distribution does not show specific trends with $n$ on the cluster-scale and galaxy-scale dPIEs. Most parameters agree within $3\sigma$ among all models on a specific set. $r_{\rm cut}$ estimates related to BCGs present a statistically significant dispersion between models, but they are the only estimations with these properties. The estimations obtained for the scaling relation parameters, $\sigma_0^*$ and $r_{\rm cut}^*$ are shown in Fig.~\ref{fig:sigposVSrcut} for models optimised on both sets of constraints, the uncertainties represent $68$ per cent of the distribution. As we can see, there are no striking changes between models if we except $\sigma_0^*$ estimation of the model with $n=3$ on the first set of constraints which shows significantly lower values compared to other models. Thus, it is the only main difference that is seen among all dPIE parameters. This is due to the large standard deviation of their prior as these parameters are less constrained compared to the BCGs or DM clumps parameters. Similarly, no particular trend is observed for the perturbation centre, the position angle, and $d_{\rm latt}$. Estimations of these parameters converge to different values that are uniformly spaced within the prior distribution. Only a few of these estimations are in agreement within $3\sigma$ between models.

\subsubsection{2D convergence field}
\begin{figure*}
\centering 
\includegraphics[width=\linewidth]{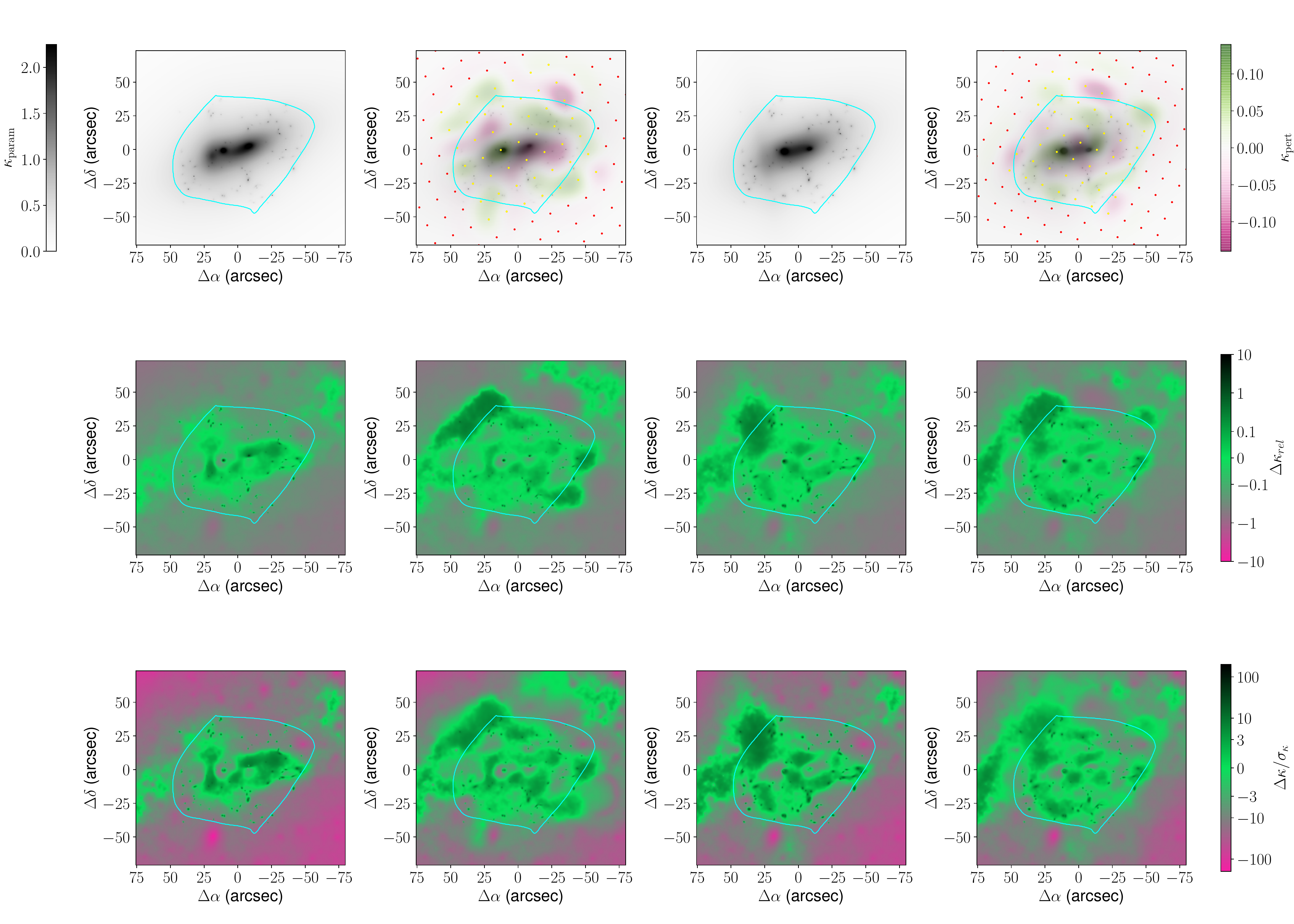}
\caption{Results obtained on the convergence maps. Top row, in reading order: maps of $\kappa$ for the fiducial model and the perturbed one with $n=7$ optimised on the first set of multiple images, and the same maps for the second set of constraints. For perturbed models, the convergence due to the \textit{parametric} component is shown in grey scale, while the \textit{free-form} component follows a colour scheme where green is positive and pink is negative. The cyan closed-curve delimits the constrained area; it represents the area that can contain multiple images for sources at $z=7$. For perturbed models, the position of knots of the B-spline mesh are represented by red dots, the ones that are coincident with non-zero basis functions are in green and follow the scheme introduced in Fig.~\ref{fig:meshknot}. Second and third row: same as the first row for $\Delta\kappa_{\rm rel}$ maps and $\Delta\kappa/\sigma_\kappa)$ maps, respectively.}
\label{fig:mass_distrib}
\end{figure*}

The top row of Fig.~\ref{fig:mass_distrib} shows the convergence $\kappa$ obtained for the best models with and without perturbation ($n=7$), for both set of constraints. The distribution due to the \textit{parametric} component is shown in grey scale while the change in $\kappa$ caused by the perturbation is shown in colour depending on whether it is increasing (green) or decreasing (pink) $\kappa$. The convergence $\kappa_{\rm pert}$ due to the perturbative modelling shows a good agreement between both sets of constraints, the pink and green regions are similar.

We can see that for the first set of constraints on the left column of Fig.~\ref{fig:mass_distrib}, the perturbation has a more significant contribution compared to the second set of constraints. Especially, the standard deviation of pixel values inside a disk centred on the cluster centre with a radius of $100$~arcsec is $20$ per cent higher and the peak-to-valley amplitude is $10$ per cent higher than the model fitted on the second set. This could be because the fiducial model used to define the dPIE parameter priors has a poorer reproduction of multiple images resulting in a $\sim50$ per cent higher RMS. However, both perturbed models achieve a similar RMS ($0.23$~arcsec and $0.20$~arcsec for the first and the second set, respectively).

For the two sets of multiple images and for both the fiducial model and the perturbed one, the middle row of Fig.~\ref{fig:mass_distrib} presents maps of relative errors on $\kappa$:
\begin{equation}
 \Delta\kappa_{\rm rel}=(\kappa_{\rm best}-\kappa_{\rm truth})/\kappa_{\rm truth}
\end{equation}
where $\kappa_{\rm best}$ is the convergence from the best-fitting model and $\kappa_{\rm truth}$ is the convergence from the simulation. The bottom row of Fig.~\ref{fig:mass_distrib} shows maps of the errors expressed in terms of $\sigma_\kappa$, the statistical uncertainties on $\kappa_{\rm best}$ obtained from the MCMC run:
\begin{equation}
    \Delta\kappa/\sigma_\kappa=(\kappa_{\rm best}-\kappa_{\rm truth})/\sigma_\kappa
\end{equation}
Adding the perturbation brings the relative error on the convergence in the range of $\pm10$ per cent over a larger area. However, this comes at the cost of a larger relative error near the limits of the perturbation where there are fewer constraints. This is especially visible near $\Delta \alpha\sim-40$~arcsec and $\Delta \delta\sim40$~arcsec, where there is a hole in the mass distribution. A similar pattern can be seen at $\Delta \alpha\sim-60$~arcsec and $\Delta \delta\sim-25$~arcsec for models on the first set and at $\Delta \alpha\sim-25$~arcsec and $\Delta \delta\sim-50$~arcsec for the second.
Nevertheless, according to the last row in Fig.~\ref{fig:mass_distrib}, the ratio of the error and the statistical uncertainties are not increased in these regions compared to the fiducial, it is the opposite. Even if the best-fitting model with $n=7$ can have a worse reproduction of these area due to fewer information provided by the constraints, it is accounted for in the statistical error. The same behaviour is exhibited but with a peak of mass instead, along the upper-left edge of the B-splines surfaces. In the same way, larger statistical uncertainties are propagated at these positions.

\subsubsection{Errors on the 2D convergence field}
\begin{figure*}
    \begin{minipage}{0.49\linewidth}
    \centering
    \includegraphics[width=\linewidth]{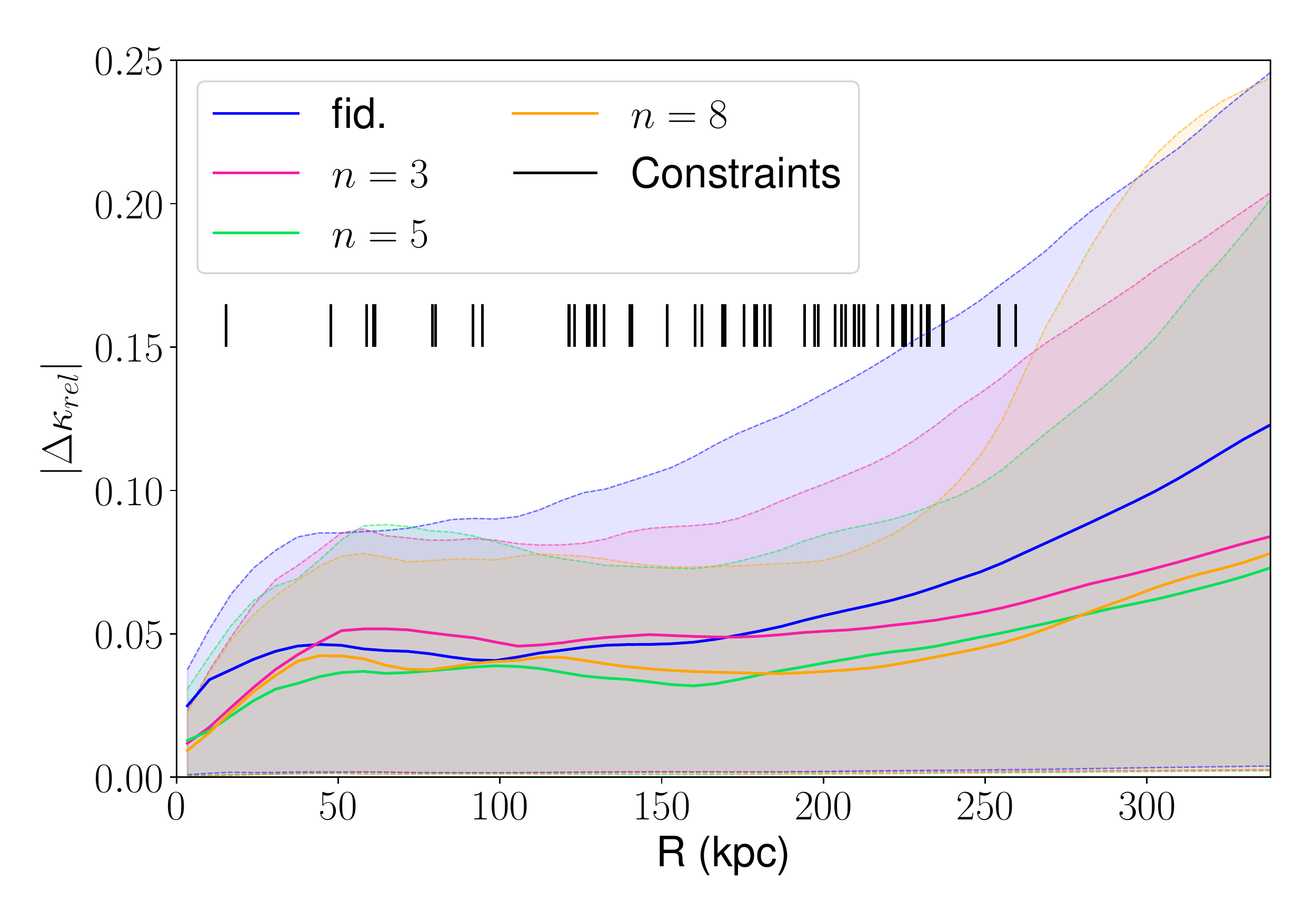}
    \end{minipage}
    \begin{minipage}{0.49\linewidth}
    \centering
    \includegraphics[width=\linewidth]{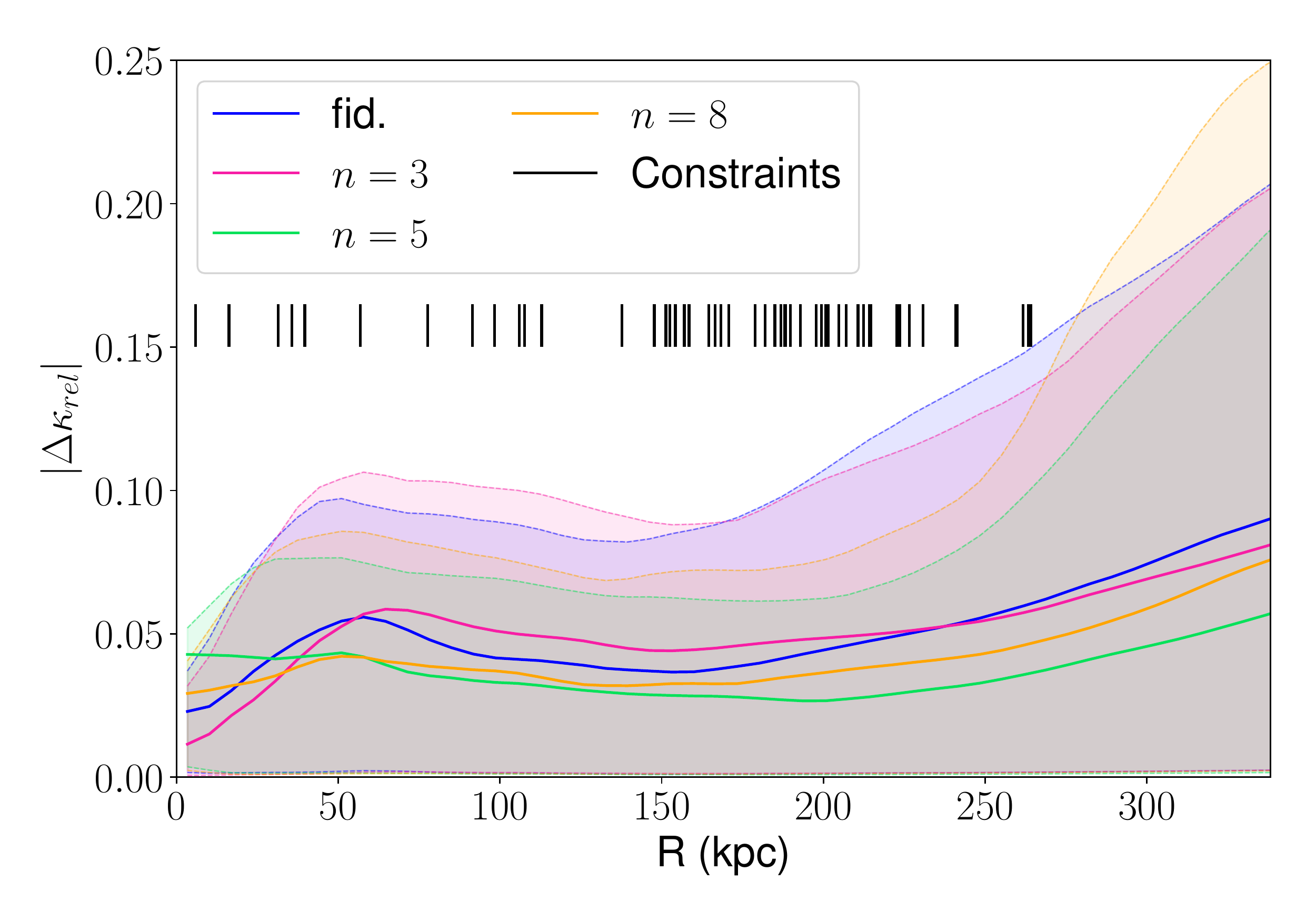}
    \end{minipage}
    
    \caption{\textit{Left panel}: the median $|\Delta\kappa_{\rm rel}|$ measured in an enclosed circle centred on the cluster centre. It shows how accurately the best models reproduce the correct data for the first set of constraints. The error shows $68$ per cent of the distribution. \textit{Right panel}: same plot for the second set of constraints.}
    \label{fig:err_relat}
\end{figure*}

To better quantify the impact of the perturbation, Fig.~\ref{fig:err_relat} presents the distribution of $|\Delta\kappa_{\rm rel}|$ contained inside a circle of growing radius and centred on the cluster centre (e.g. Fig.~\ref{fig:kappa_sim} left panel). The shaded areas show $68$ per cent of the distribution while the central lines represent medians, and the small black bars show the radius of multiple images relative to the cluster centre. For more clarity, only four models are plotted for each set of constraints to illustrate the trends.

Perturbed models with $n\geq5$ show an improvements in the median values compared to the fiducial one for both sets of constraints. If we look at the models plotted, these perturbed models have a median lower than at least $0.5$ per cent at $R\sim60$~kpc to more than $4.5$ per cent at the edge of the constrained area for the first set of constraints. The improvement is smaller for the second set of constraints, there median values are only lower of at least $1.5$ per cent for $R\geq280$~kpc. Most notably, the fiducial model associated with the second set of constraints outperforms the one on the first set. Only models with $n=5$ show a similar behaviour with a better result on the second set, when other perturbed models present similar $|\Delta\kappa_{\rm rel}|$ between their optimisation on the different realisation of constraints. Thus, this sensitivity to the realisation of constraints explains why the difference on the median values between perturbed and fiducial models is smaller for the second set. This also shows a weaker dependency on a specific set of constraints for perturbed models compared to fiducial ones.

The best improvements of perturbed models compared to fiducial ones are on the upper bound at $84$ per cent. These enhancements are present on models with $n\geq5$ in the constrained area (e.g. $R\leq280$~kpc). The differences with the fiducial models are the largest in the range $130$~kpc~$<R<280$~kpc where the upper bound for perturbed models are almost constant and below $10$ per cent. The same limit is exceeded for fiducial models at $R\sim150$~kpc and $R\sim210$~kpc for the first and second set, respectively. Hence, most perturbed models tend to reduce the width of the error on the convergence field compared to the fiducial ones in the constrained area as it is shown qualitatively in Fig.~\ref{fig:mass_distrib} and by the median and upper bound values.

\begin{figure*}
    \begin{minipage}{0.49\linewidth}
    \centering
    \includegraphics[width=\linewidth]{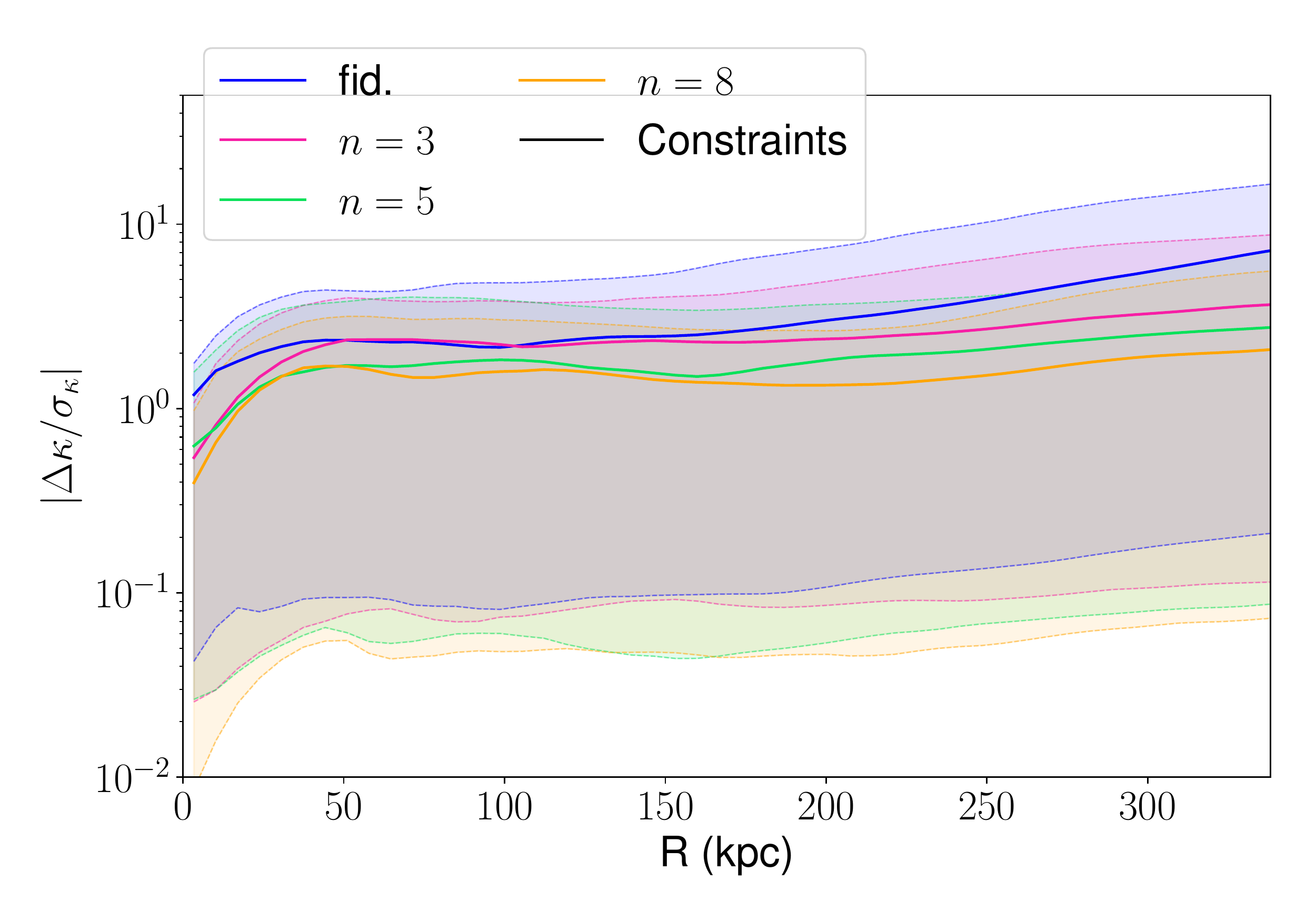}
    \end{minipage}
    \begin{minipage}{0.49\linewidth}
    \centering
    \includegraphics[width=\linewidth]{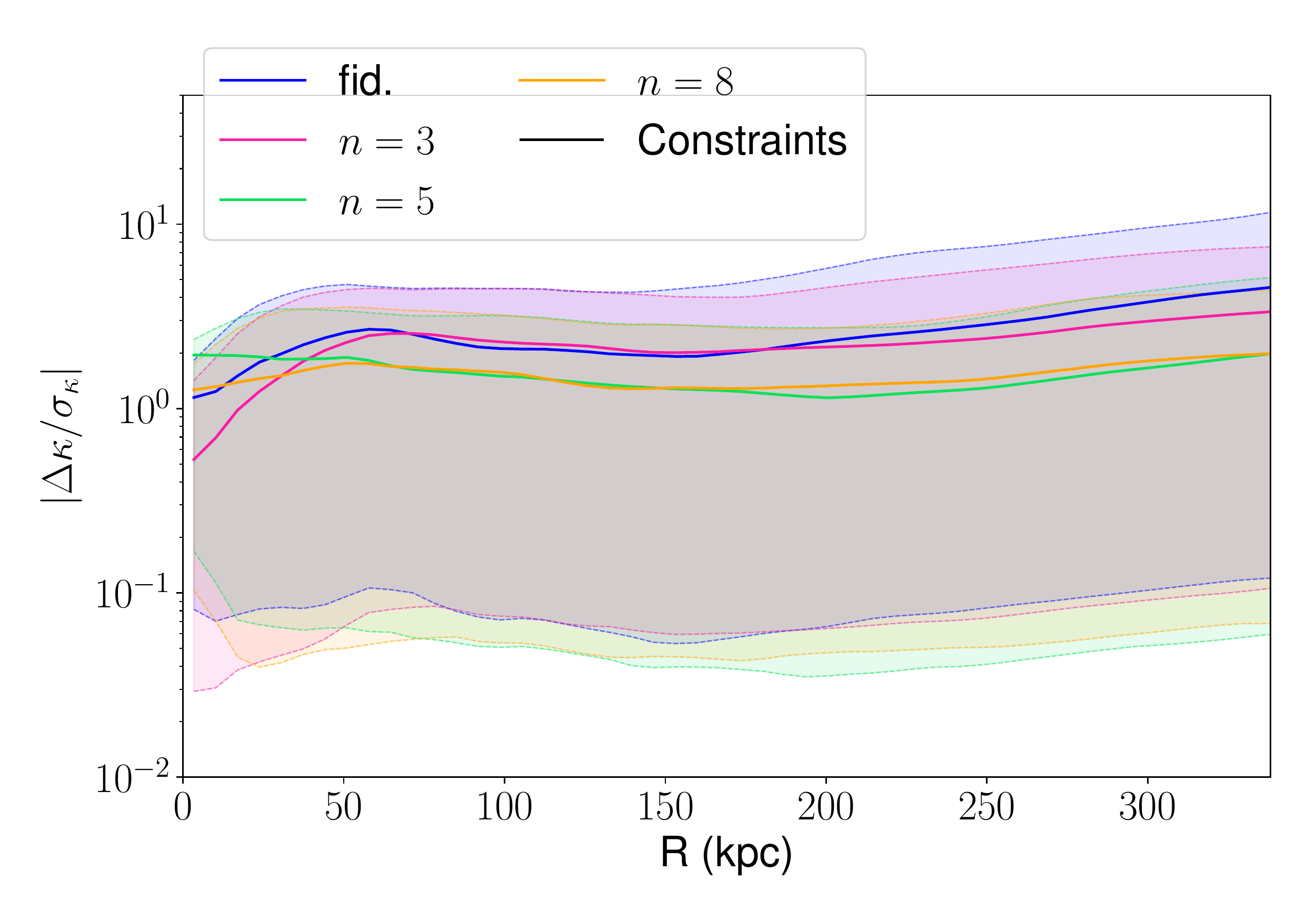}
    \end{minipage}
   
    \caption{Same as Fig.~\ref{fig:err_relat} but for $|\Delta\kappa/\sigma_\kappa|$ maps.}
    \label{fig:err_chi}
\end{figure*}

Similarly to Fig.~\ref{fig:err_relat}, Fig.~\ref{fig:err_chi} shows the distribution of $|\Delta\kappa/\sigma_\kappa|$ inside a circle of growing radius. As the results on $|\Delta\kappa_{\rm rel}|$, the median values for fiducial and perturbed models with $n=3$ and $n=4$ are the highest ones for both set of constraints. Most notably, for perturbed models with $n>4$, the median values are constantly within $3\sigma$ and $2\sigma$ from the simulated data for $R<340$~kpc for models on the first set and on the second set, respectively. If we look at the upper bound at $84$ per cent, they show an agreement at less or similar to $7\sigma$ and $5\sigma$ for the same models and the same radius. For the fiducial models, the median values are contained between $1\sigma$ to $7\sigma$ when the upper bound is between $2\sigma$ to $16\sigma$ for both set of constraints. Hence, as we have seen in Fig.~\ref{fig:err_relat}, median values are better for most perturbed models compared to the fiducial and the best improvements are especially seen on the upper bound at $84$ per cent. These enhancements show that the addition of the perturbative surface makes the estimation of the errors on the convergence through the statistical uncertainties much more accurate. It constitutes evidence that the perturbation allows for more flexibility on the mass reconstruction, which could vary more homogeneously than with the dPIEs only. Particularly, only a few multiple images can well constrain a dPIE in areas where no observables are accessible. In the case of a perturbative surface, when there are no constraints in a region, the associated statistical uncertainties are increased.

\subsection{Radial profiles}
\label{sec:profiles}
\begin{figure*}
    \begin{minipage}{0.49\linewidth}
    \centering
    \includegraphics[width=\linewidth]{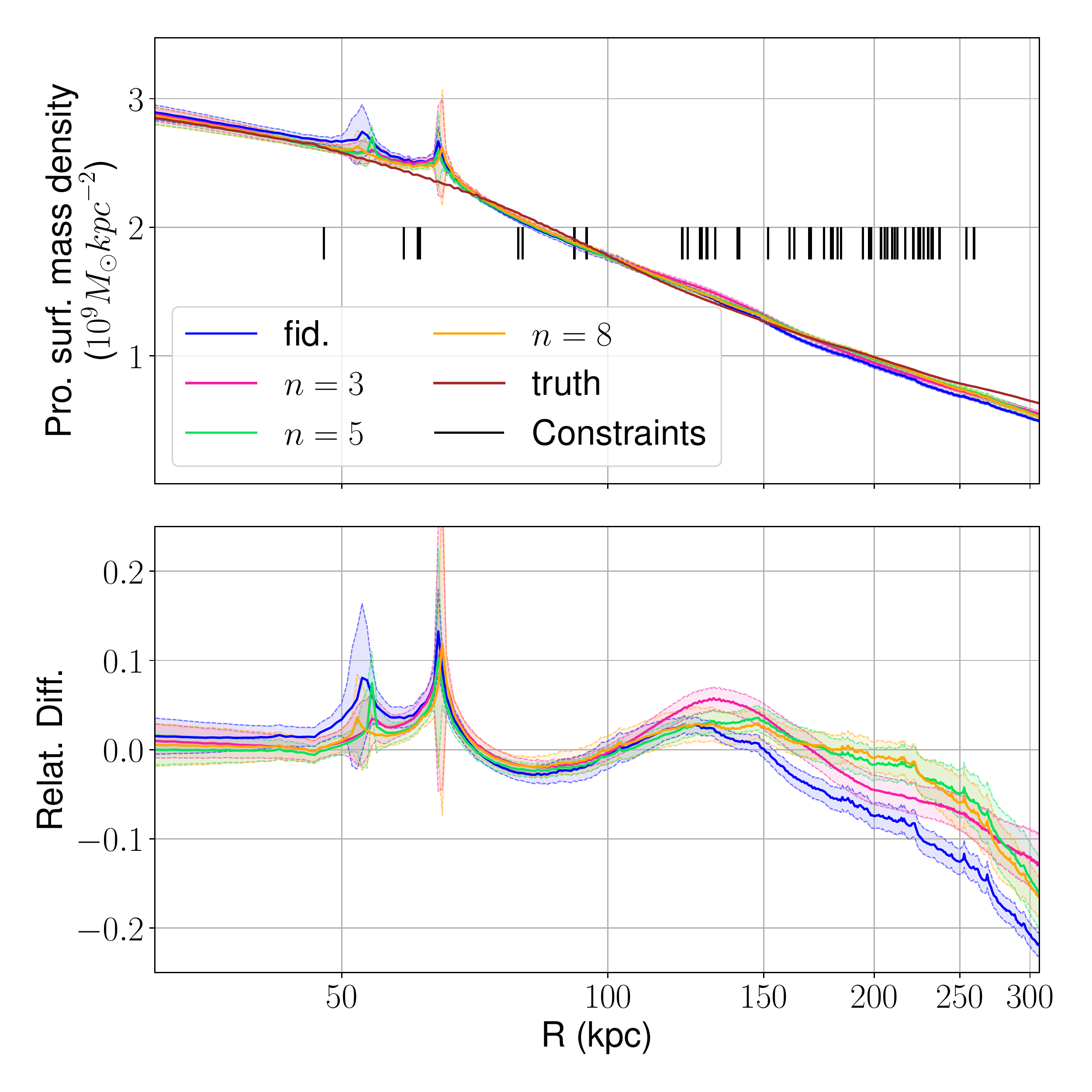}
    \end{minipage}
    \begin{minipage}{0.49\linewidth}
    \centering
    \includegraphics[width=\linewidth]{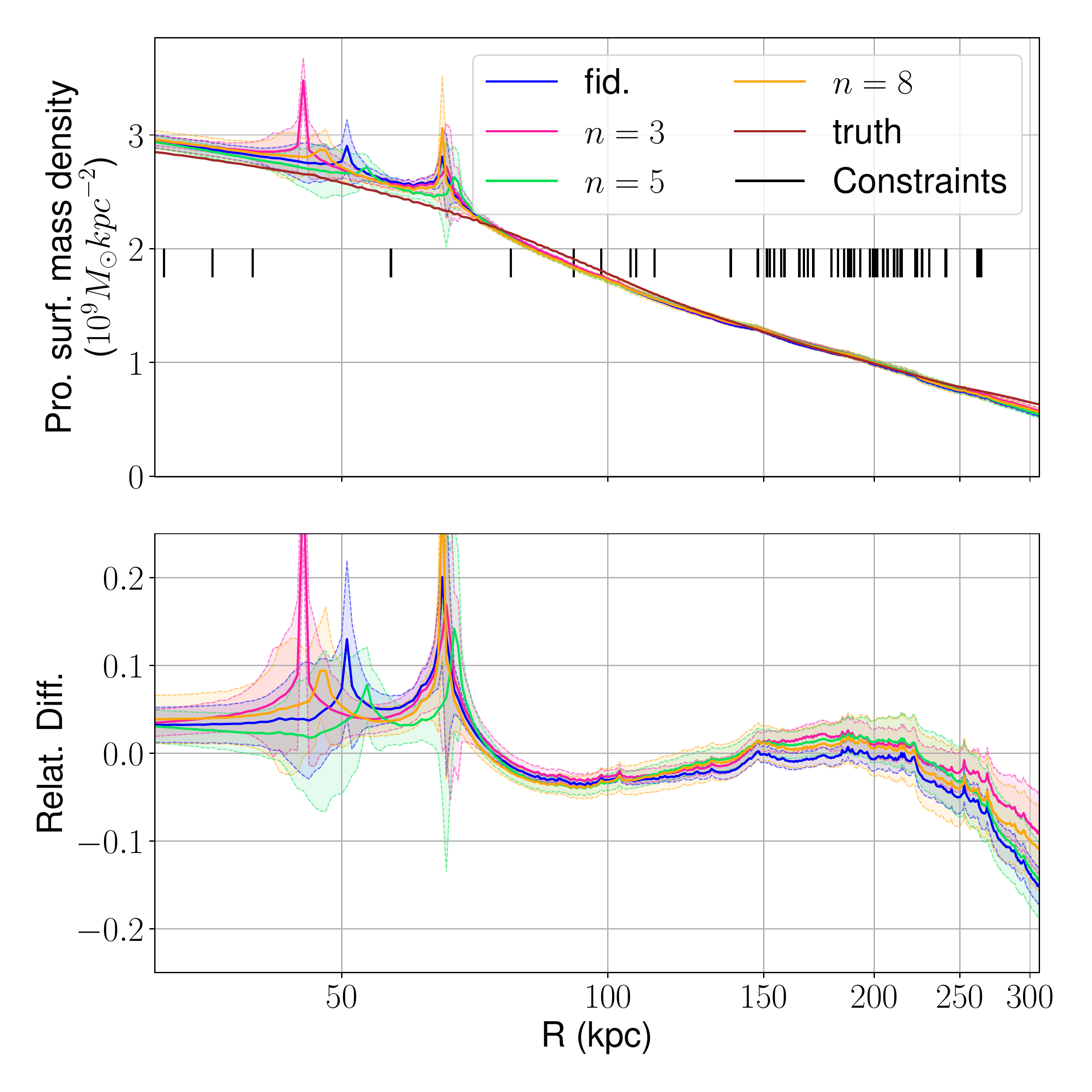}
    \end{minipage}
    
    \caption{Radial profiles of the mass distribution. First row, profiles computed with the mean surface density enclosed in concentric annuli around the cluster centre for best-fitting models and the true mass distribution. The black bars represent the distance of the multiple images from the cluster centre. Profiles on the left panel and on the right panel are associated with models optimised on the first set and on the second set of constraints, respectively. Second row, the relative difference between profiles from best-fitting models and the true distribution. In both rows, the errors represent three times the statistical uncertainties from the MCMC runs.}
    \label{fig:circular_prof}
\end{figure*}

We investigate the reproduction of radial mass profiles by computing the mean surface density enclosed in concentric annuli around the cluster centre shown in Fig.~\ref{fig:kappa_sim}. These profiles are shown in Fig.~\ref{fig:circular_prof} for the best models. The shaded areas represent three times the statistical uncertainties and the black ticks indicates the multiple images distances with respect to the centre.

Following the results on the convergence field discussed in Section~\ref{sec:mass_dist}, the true mass profile is not well-recovered at large radii by the fiducial model obtained with the first set of constraints. But even in this case, a perturbation with $n\geq5$ appears to be effective at improving the profile. All other models agree with each other at the $3\sigma$ level for $R<600$~kpc, which is roughly the limit of the publicly available data on \textit{Hera}. Overall, ignoring the peaks due to the BCGs at $R\sim60$~kpc, those models recover the true profile to within $5$ per cent inside the constrained area ($R\leq250$~kpc) but consistently underestimate the mean surface density by $\sim10\%-35$ per cent outside. 

Although the improvement on the radial profile offered by the use of a perturbation is only marginal, it is able to partially correct the profile when the \textit{parametric} modelling initially under-performs. We note however that an accurate recovery of the profile on large scales depends partly on the distribution of multiple images, as illustrated by the small differences between the two sets.

\subsection{Effective Einstein radii and critical lines}
\label{sec:Einstein_radii}
\begin{figure*}
    \begin{minipage}{0.49\linewidth}
    \centering
    \includegraphics[width=\linewidth]{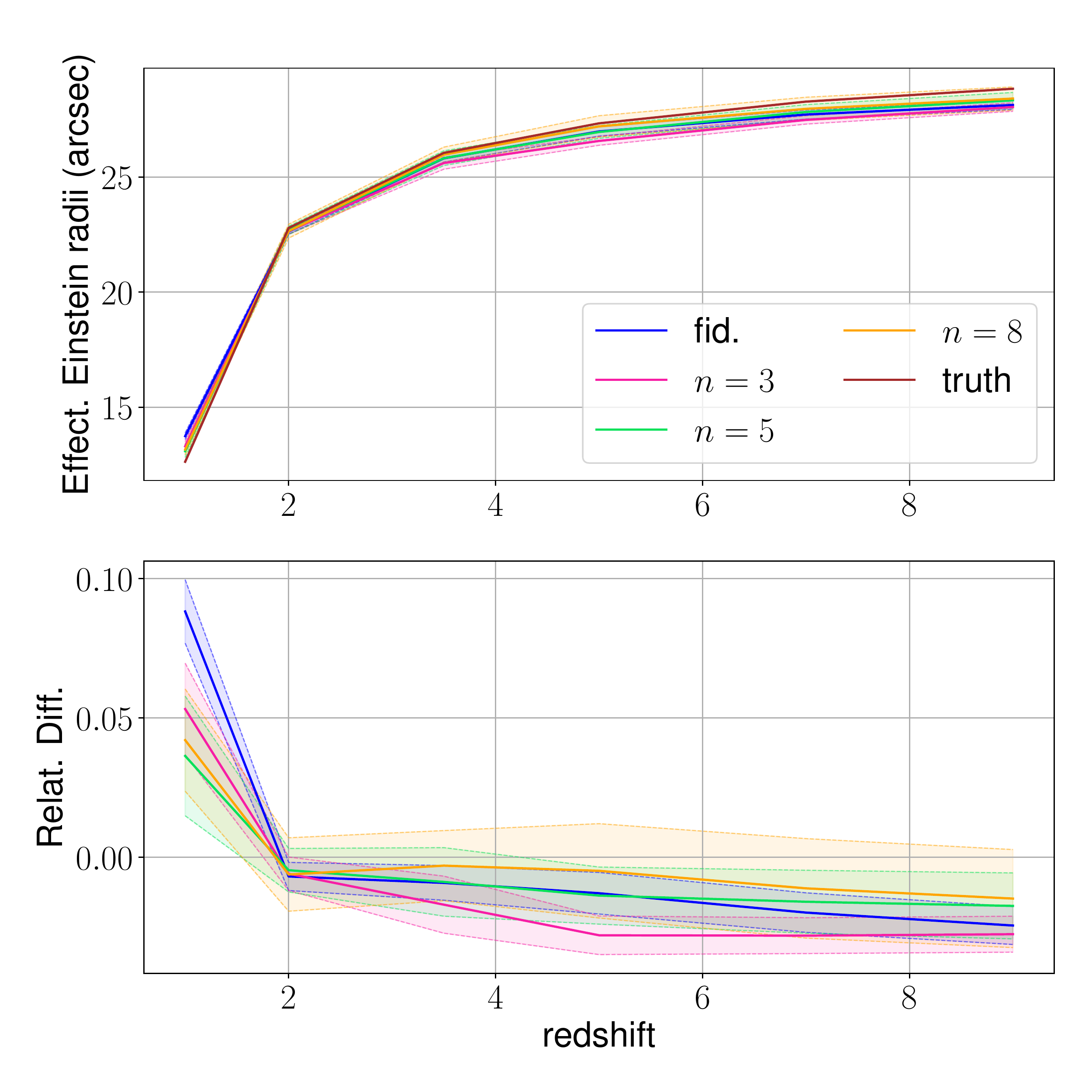}
    \end{minipage}
    \begin{minipage}{0.49\linewidth}
    \centering
    \includegraphics[width=\linewidth]{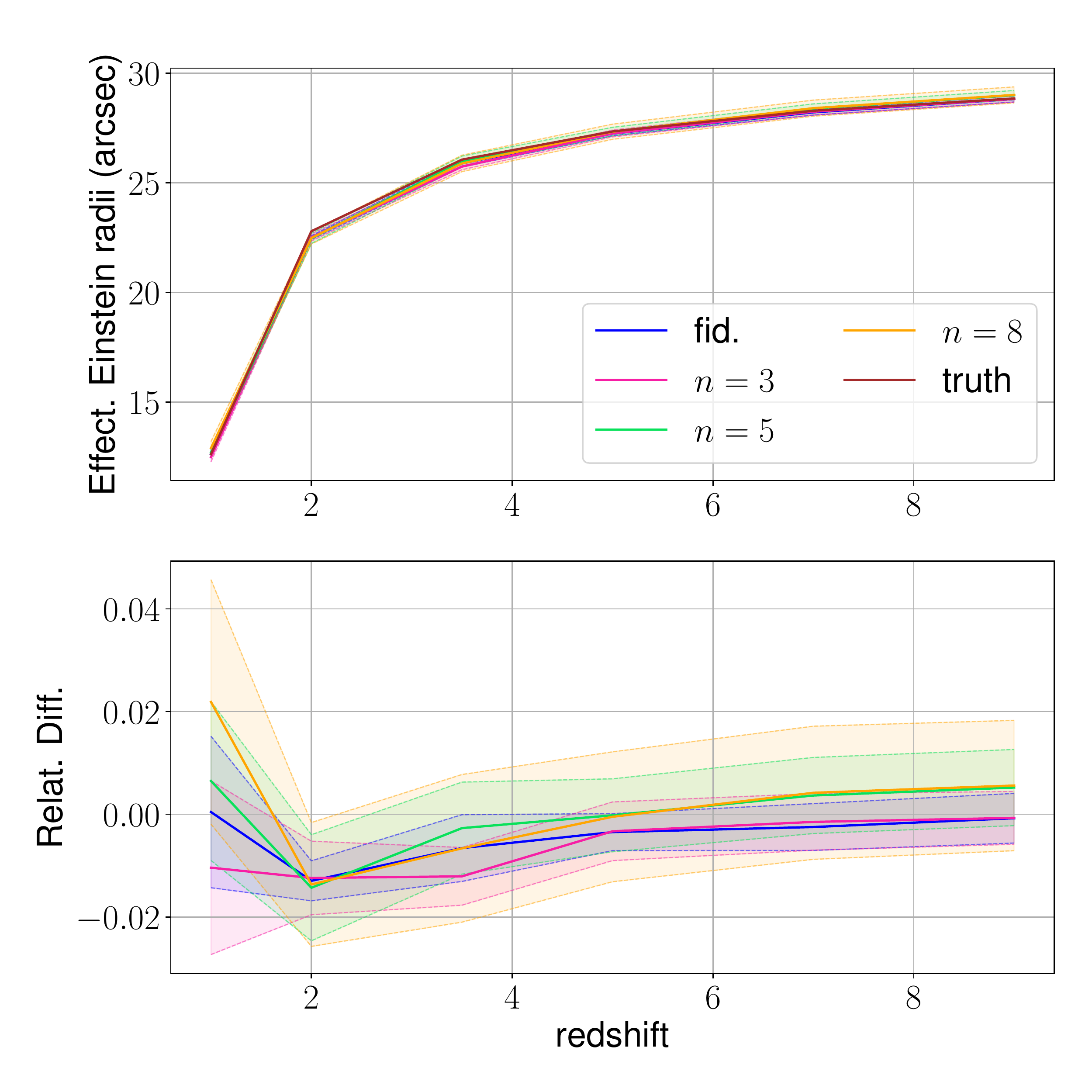}
    \end{minipage}
   
    \caption{\textit{Left panel}: Effective Einstein radii. First row, Einstein radii for best-fitting models optimised on the first set of constraints and the simulated data of Hera cluster. Second row, relative difference between the radii from the best-fitting models and the true data. For both rows, the shaded areas represent three times the statistical uncertainties. \textit{Right Panel}: same plot for the second set of multiple images.}
    \label{fig:einstein_radii}
\end{figure*}

Following \citet{Redlich2012}, we define the \textit{effective} Einstein radius from the surface inside critical lines $S_{\rm cl}$ at a given redshift :
\begin{equation}
    \theta_E \text{(arcsec)}=\sqrt{\frac{S_{\rm cl}}{\pi}}
    \label{eq:einstein_radii}
\end{equation}
Fig.~\ref{fig:einstein_radii} shows Einstein radii computed between $z=1$ and $z=9$ for the best models and their relative difference with respect to the simulated data. The shaded area represents three times the statistical uncertainties from the MCMC run.

\begin{figure*}
    \begin{minipage}{0.49\linewidth}
    \centering
    \includegraphics[width=\linewidth]{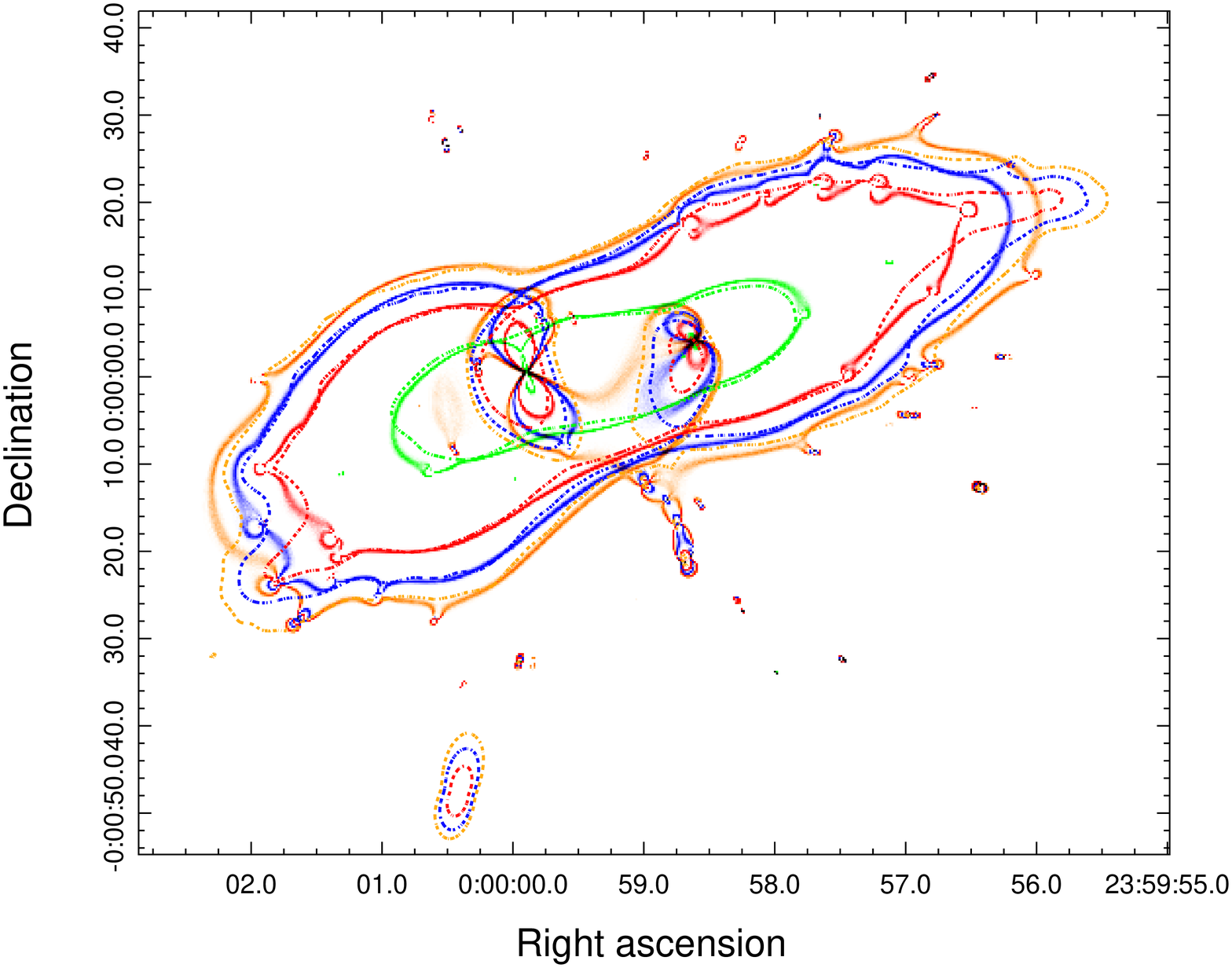}
    \end{minipage}
    \begin{minipage}{0.49\linewidth}
    \centering
    \includegraphics[width=\linewidth]{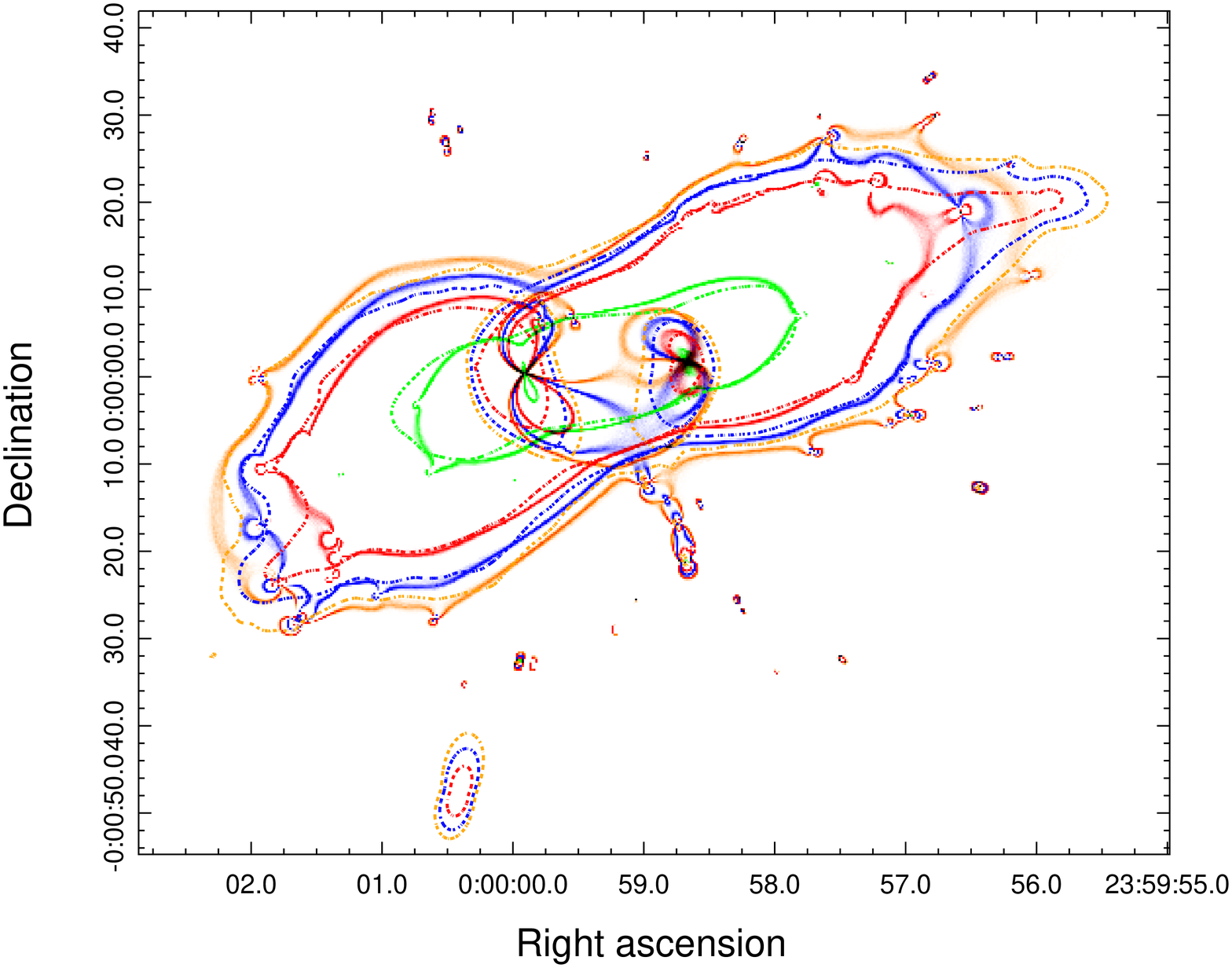}
    \end{minipage}
    \caption{Critical lines from 1000 perturbed models with $n=6$ from the output of the MCMC chains along with the ones from the simulated data of \textit{Hera} cluster. The lines are computed for $z=1$ (green), $2$ (red), $3.5$ (blue) and $7$ (orange), solid and dotter lines are associated with the perturbed models and the true data, respectively. Models on the left and right panel are fitted on the first and second sets of multiple images, respectively.}
    \label{fig:critic_line}
\end{figure*}

All the models agree to within $3\sigma$ for both sets of constraints. They agree within $3\sigma$ with the truth for most of the computed redshifts, with relative differences being mostly below $5$ per cent. For the first set of images, Einstein radii for $z=1$ and $z\geq5$ are slightly more than $3\sigma$ away from the truth while only the Einstein radius for $z=2$ is slightly underestimated for the second set. However, we note that the critical lines are less extended to the right side of the cluster in the models compared to the simulation. This is due to a galaxy that is not modelled by a dPIE profile in the model provided for the HFF challenges\citep{meneghetti2017}. Ignoring the last fact, the tangential critical lines from perturbed models are close to the true ones as we can see in Fig.~\ref{fig:critic_line} which shows the critical lines for 1000 models from the sample with $n=6$ for both set of constraints. However, the radial critical lines are not well reproduced and show significant differences with the one from the simulation. We can also see that the distribution of these lines is widening as they are further away from the constrained area. This behaviour is more and more significant as $n$ is increasing.

dPIE profiles used to model cluster members have a more concentrated distribution at their centre in comparison with the simulated data, which result in more critical lines for the considered redshifts on the galaxy-scale.Thus, this can compensate for the less extended critical lines on the right side and the galaxy that is not modelled in computations of the effective Einstein radii. \citet{Meneghetti2020} showed that a similar behaviour is even present in more recent state-of-the-art simulations. They found that cluster members in simulated clusters tend to produce less secondary critical lines compared to observations. Overall, as all of these models show similar patterns in reproducing the critical lines, the addition of a perturbation does not significantly improve the estimation of the effective Einstein radii.

\subsection{Substructure masses}
\label{sec:mass_sub}

\begin{figure*}
    \begin{minipage}{0.49\linewidth}
    \centering
    \includegraphics[width=\linewidth]{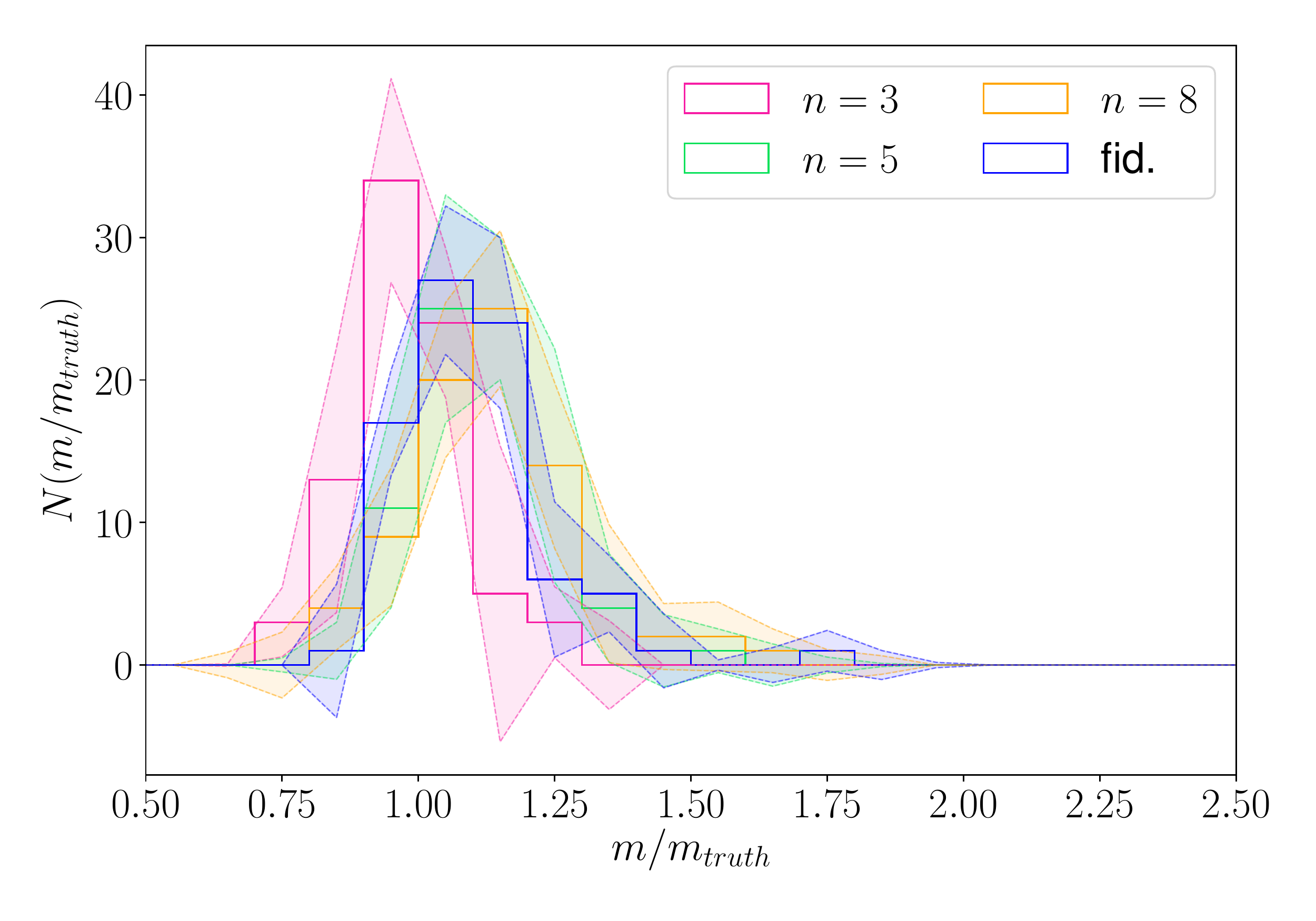}
    \end{minipage}
    \begin{minipage}{0.49\linewidth}
    \centering
    \includegraphics[width=\linewidth]{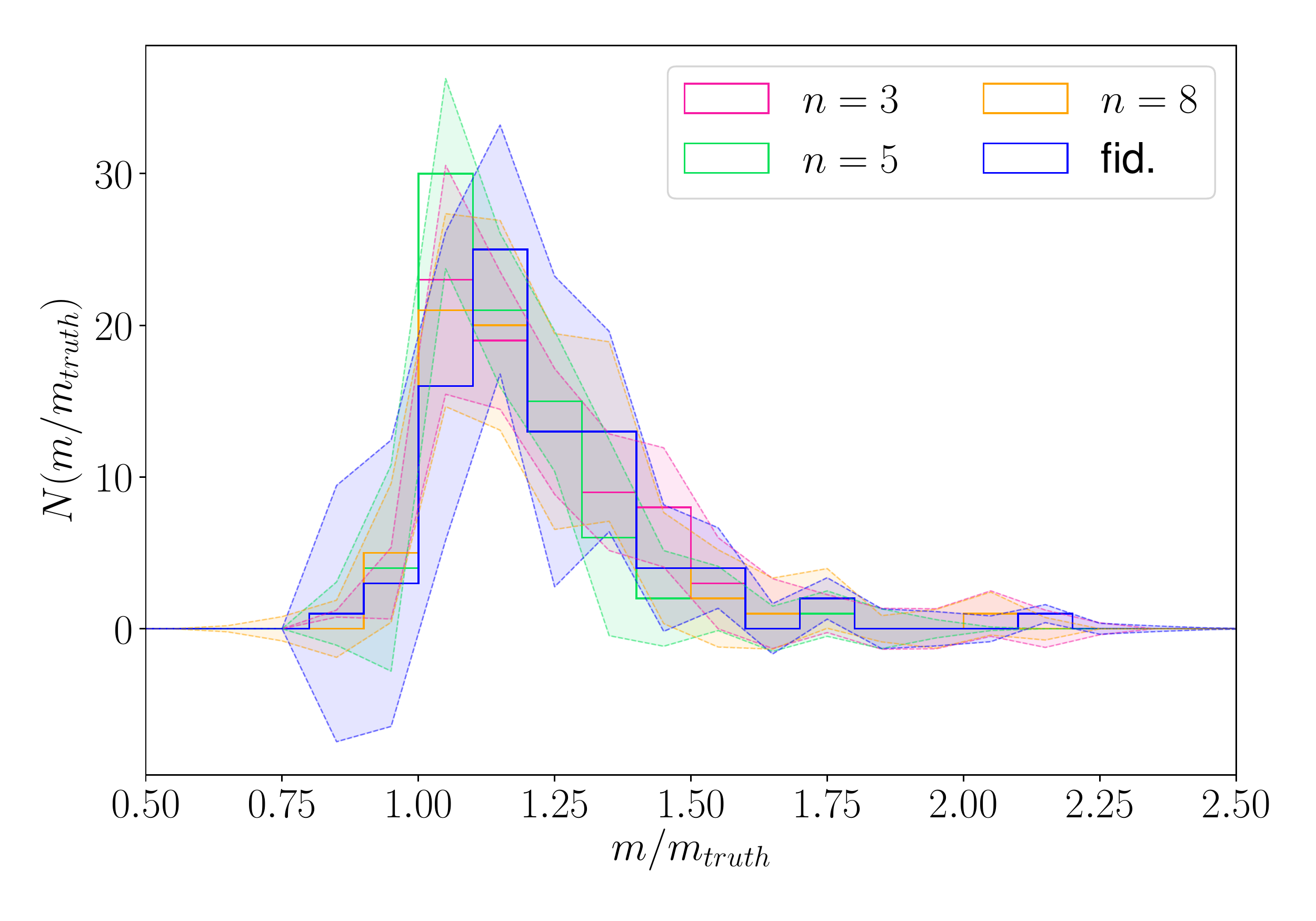}
    \end{minipage}
    
    \caption{Histograms of the distribution of the ratio between the mass of cluster members computed on the best-fitting models and on the simulated data. These masses are computed as the mass enclosed in a circle of radius $r_{\rm eff}$ centred on the centroid of the associated galaxy luminosity distribution. The shaded areas represent the statistical uncertainties from the output of the MCMC. Histograms on the left panel and on the right panel are associated with models optimised on the first set of multiple images and on the second one, respectively.}
    \label{fig:substruct}
\end{figure*}

We investigate the reproduction of the galaxy-scale components by comparing the mass enclosed in a circle around cluster members in our models and in the simulation. We only consider galaxies that are modelled by a dPIE profile, because even if our perturbation increases the model flexibility, its scale is still much larger than the spatial scale of the cluster members. To be more specific, if we compare the median distance between $r_{\rm core}$ and $r_{\rm cut}$ for cluster members, excluding BCGs, and the distance between two B-spline bases $d_{\rm latt}/(n + 1)$, the first ranges between $1$~arcsec to $10$~arcsec when the second ranges between $8$~arcsec ($n=12$) to $30$~arcsec ($n=3$). The effective radius used to compute the mass is defined as $r_{\rm eff}=10\sqrt{ab}$ where $a$ and $b$ are the semi-major axis and semi-minor axis of an ellipse describing the galaxy light distribution. Histograms in Fig.~\ref{fig:substruct} show the distribution of the ratio between the mass estimated by best-fitting models and the mass calculated with the true distribution, errors represent three times the statistical uncertainties.

Fig.~\ref{fig:substruct} shows that the mass distributions of galaxy-scale components are similar for all models with the exception of the model with $n=3$ for the first set of constraints. This is consistent with Fig.~\ref{fig:sigposVSrcut} where estimates of $r_{\rm cut}^*$ and $\sigma_{0}^*$ are found to be mostly in agreement except for the particular model at $n=3$ which has a significantly lower estimation of $\sigma_{0}^*$. Thus, this shows that perturbed models produce equivalent mass distribution for cluster members if the estimations of $r_{\rm cut}^*$ and $\sigma_{0}^*$ are comparable between models.

Both perturbed models and fiducial ones exhibit a bias in overestimating the mass contained in cluster members. For the first realisation of constraints, most cluster members are between $0$ per cent to $20$ per cent more massive than the true values from the simulation. This bias is stronger with the second set of constraints, even if the RMS is lower for the fiducial and perturbed models compared to the first set at a fixed $n$. Also, a non-negligible part of the cluster members has been overestimated by more than $30$ per cent, both for perturbed and fiducial models. We can also note that a part of this bias is due to the excess of mass at the centre of dPIE profiles that is absent from the \textit{Hera} true mass distribution. However, as mentioned above it is in agreement with the observations of small-scale gravitational lensing events. Thus, if we ignore the specific models with $n=3$ on the first set of multiple images and considering that the differences between models are dominated by the $r_{\rm cut}^*$ and $\sigma_{0}^*$ estimation, there is no significant improvement in the reconstruction of galaxy-scale components. 

\section{Discussion}
\label{sec:discussion}
\subsection{Model metrics definition}

To compare more quantitatively which model provides the best reproduction of the true mass distribution, we use six metrics focused on the following measurements:

\begin{itemize}
    \item 2D Mass maps in the constrained area
    \item Radial mass profiles
    \item Substructures masses
    \item Positions of barycentres used to predict multiple images on its set
    \item Errors on multiple images positions estimated with the magnification tensor. These images are different from the ones used in the optimisation
    \item Effective Einstein radii
\end{itemize}

As we aimed at modelling the mass distribution of clusters, three of the metrics (out of six) are focused on this property. Because the 2D mass maps metric is a global measure of the distance between two distributions, it can not be used alone to probe more specific contributions to these distributions such as the mass of the substructures, the cluster-scale or the non-radial components. Radial profiles account for the cluster-scale distribution by probing the global slope of the mass distribution. Substructure masses are directly measured, and the combination of the three metrics probes the performance on the non-radial part.

We also put a specific focus on the deflection angle field with two metrics as it is directly constrained by the lens equation (i.e. Eq.~\ref{eq:lens_eq}). We assess the source plane bias with the residual error on the positions of each multiply-imaged source. Then, we look into the robustness of the fit with the error on the reproduction of constraints that are not used in the optimisation. Due to the non-linearity of the lens equation, there may be no solutions for few multiple images (i.e. $\sim10$ images among $\sim200$ others), even if the different sources of each element of the observed multiply-imaged systems are close. In that case, we cannot have meaningful statistics, so we used the magnification tensor to obtain a first-order approximation of the multiple images reproduction; similarly to what has been done by \citep{oguri2010} using the \textit{GLAFIC} software. This metric is also different from the metric based on the source positions because here, we do not use information about the true sources.  

Finally, we provide some insights on how well the magnification field is reproduced with a metric on the effective Einstein radii. It allows us to probe the cluster magnification on the cluster-scale. Along with the radial profiles and substructures masses, these radii are physical quantities that would be extracted from real data to be compared with large scale simulations to constrain DM properties, for that reason we need to favour models that reproduce them the most accurately \citep{Richard2010,Natarajan2017}.

For a given physical quantity $V$ where $N$ measured values $V_i$ are obtained and $V_{\rm true}$ is the same quantity derived from the simulation, we can express the metric as the inverse of the average absolute relative error on $V$:
\begin{equation}
    \text{metric}=\frac{1}{\sum\limits^N\limits_{i=1} \frac{1}{N}\left|\frac{V_i-V_{{\rm true},i}}{V_{{\rm true},i}}\right|}
    \label{eq:metric}
\end{equation}
To give an example on how this metric works, we will consider the case of the effective Einstein radii. First we compute $N$ radii for a given model in the output of the MCMC chains. Then the same radii are computed from the simulated data of \textit{Hera} cluster, and finally we take the mean of the relative difference of each radius between the truth and the considered model as showed by equation~\ref{eq:metric}. This expression is used for all the metrics mentioned above except for the barycentre positions and the errors on multiple images. For the last two, as we compare the position of two points in a plane we use the absolute error instead of the relative error which gives:
\begin{equation}
    \text{metric}=\frac{1}{\sum\limits^{N}\limits_{i=1} \frac{1}{N}\sqrt{(V_{x,i}-V_{{\rm true},x,i})^2 +(V_{y,i}-V_{{\rm true},y,i})^2}}
\end{equation}

From the posterior distribution, we obtain a distribution of metrics computed for each individual sample of parameters. Then, we extract the median, the lower and upper bound at $68$ per cent of the distribution and the value for the best models. We normalise each median metric and the associated error by the maximum between the upper bound at $68$ per cent obtained of all types of modelling on the two sets of constraints. Hence, metrics from both set of multiple images can be compared together, and the different points are on an equivalent scale.

\subsection{Model metrics}
\label{sec:metric}

\begin{figure*}
    \begin{minipage}{0.49\linewidth}
    \centering
    \includegraphics[width=1.\linewidth]{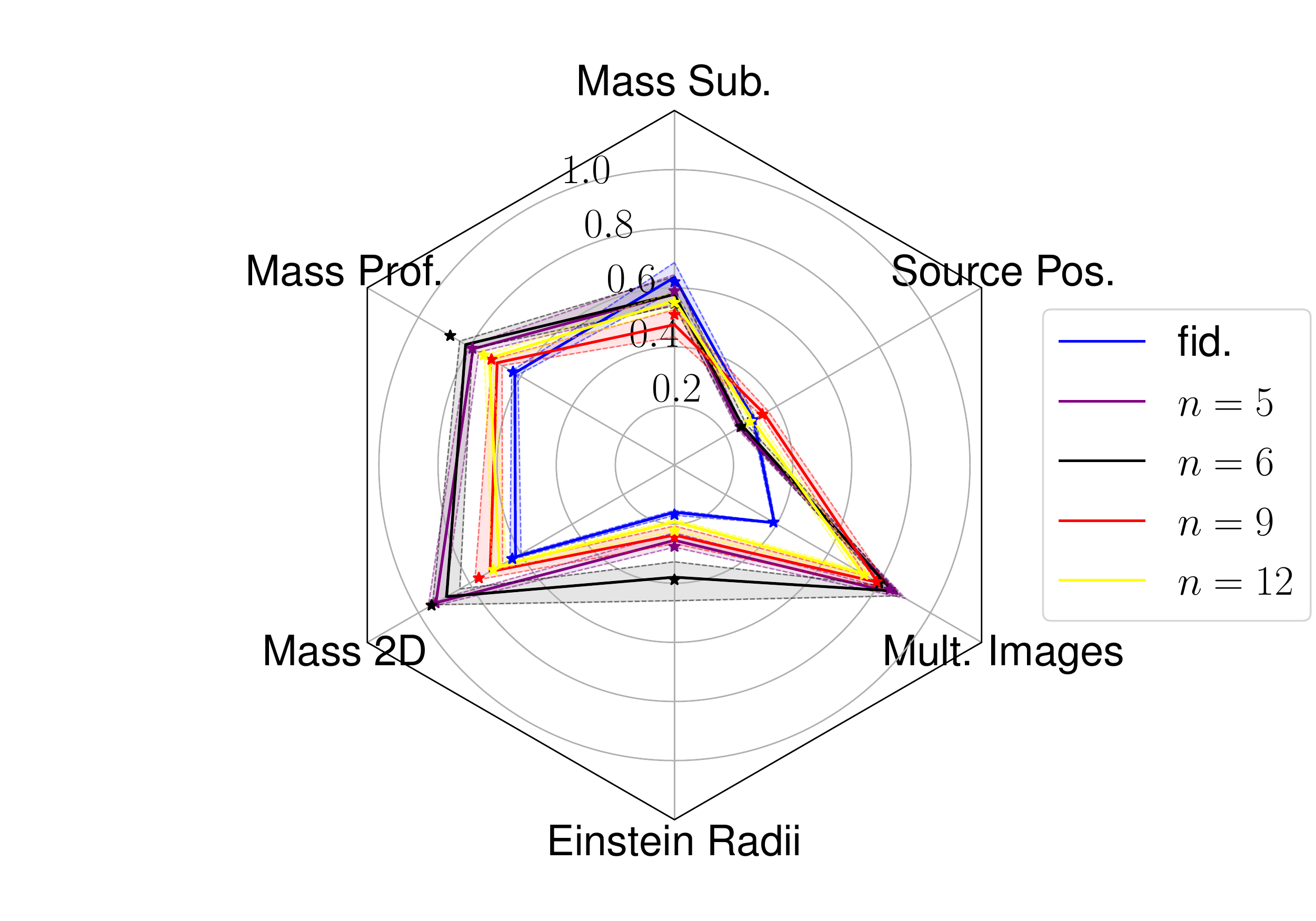}
    \end{minipage}
    \begin{minipage}{0.49\linewidth}
    \centering
    \includegraphics[width=1.\linewidth]{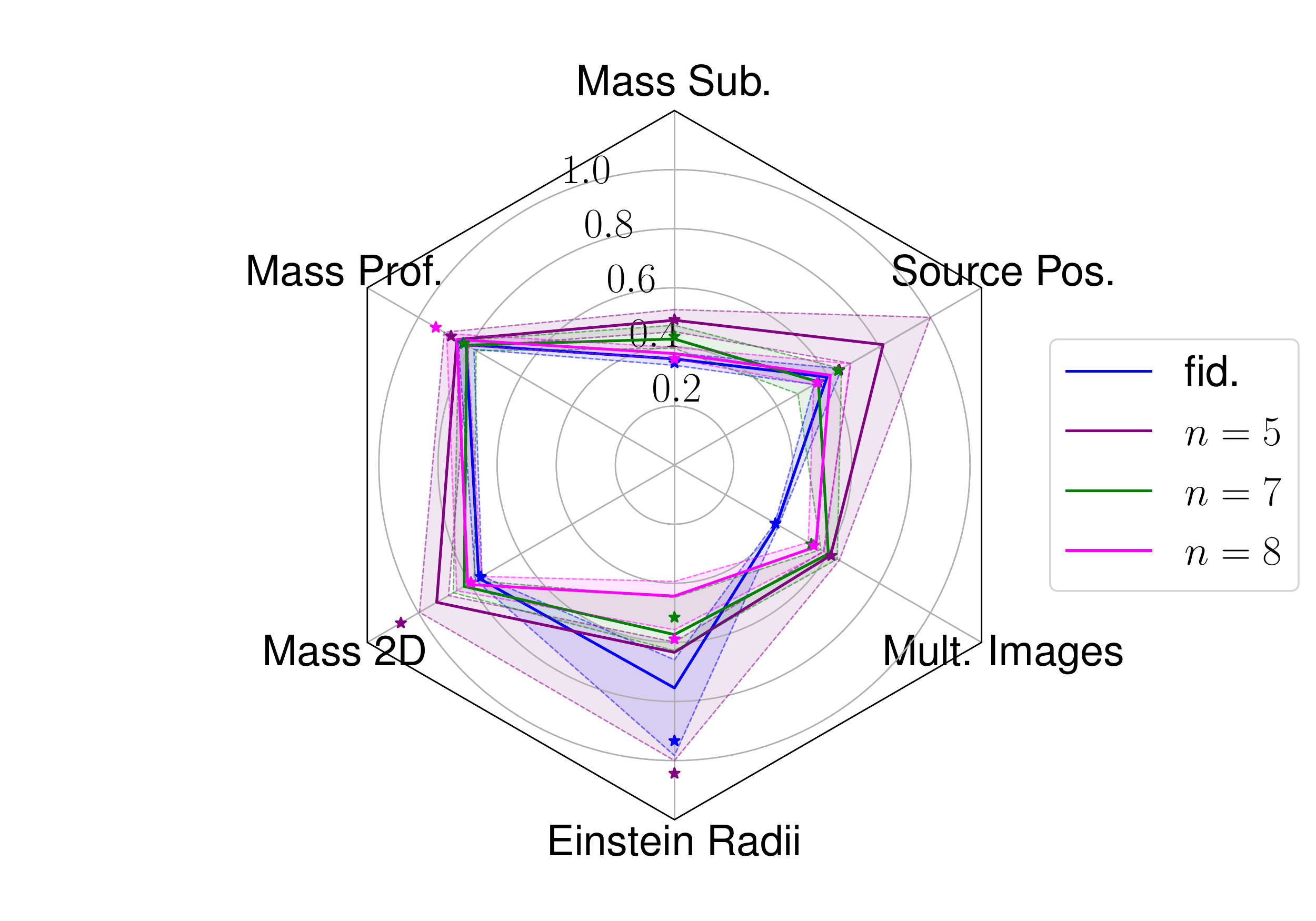}
    \end{minipage}
    \begin{minipage}{0.49\linewidth}
    \centering
    \includegraphics[width=0.8\linewidth]{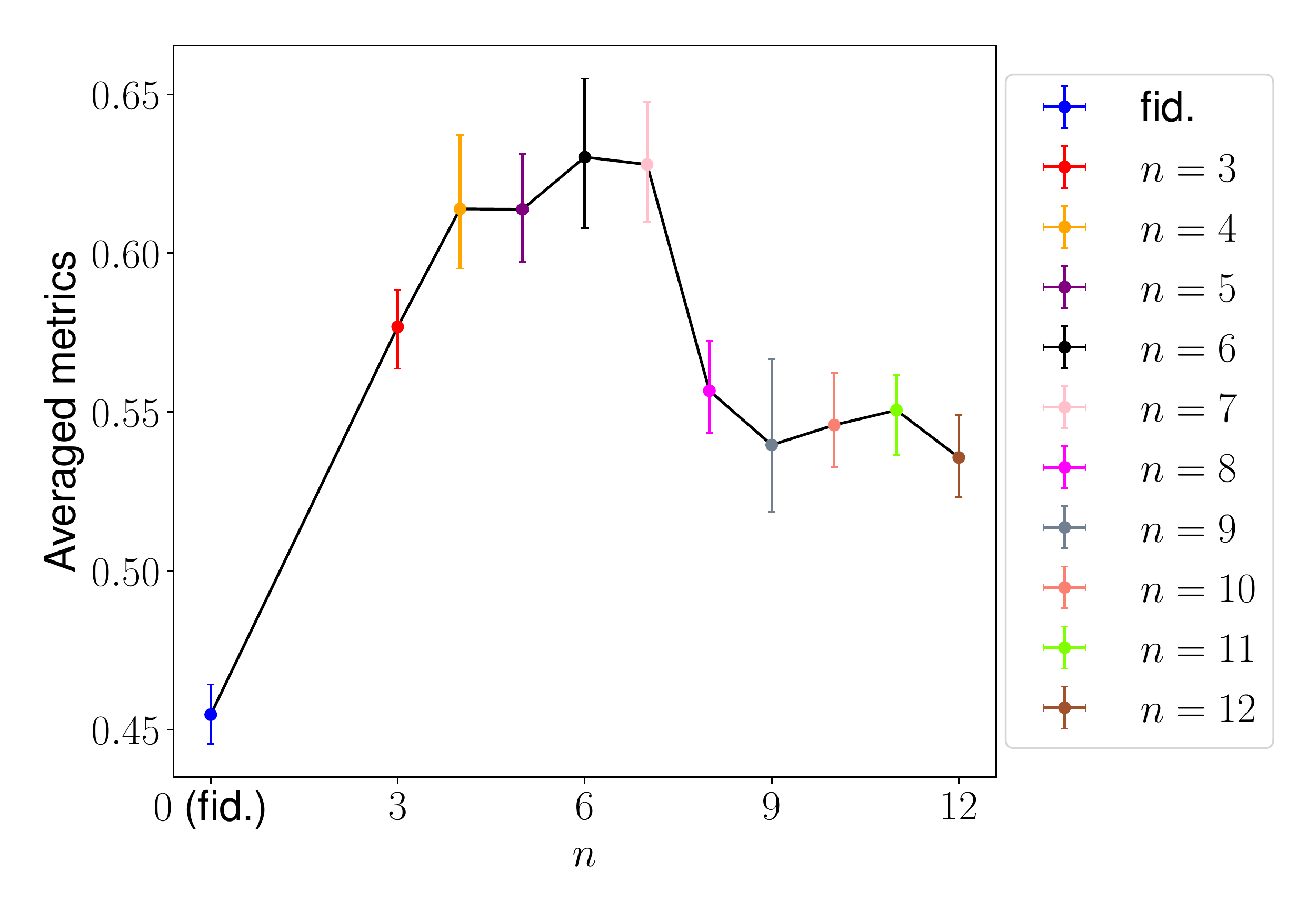}
    \end{minipage}
    \begin{minipage}{0.49\linewidth}
    \centering
    \includegraphics[width=0.8\linewidth]{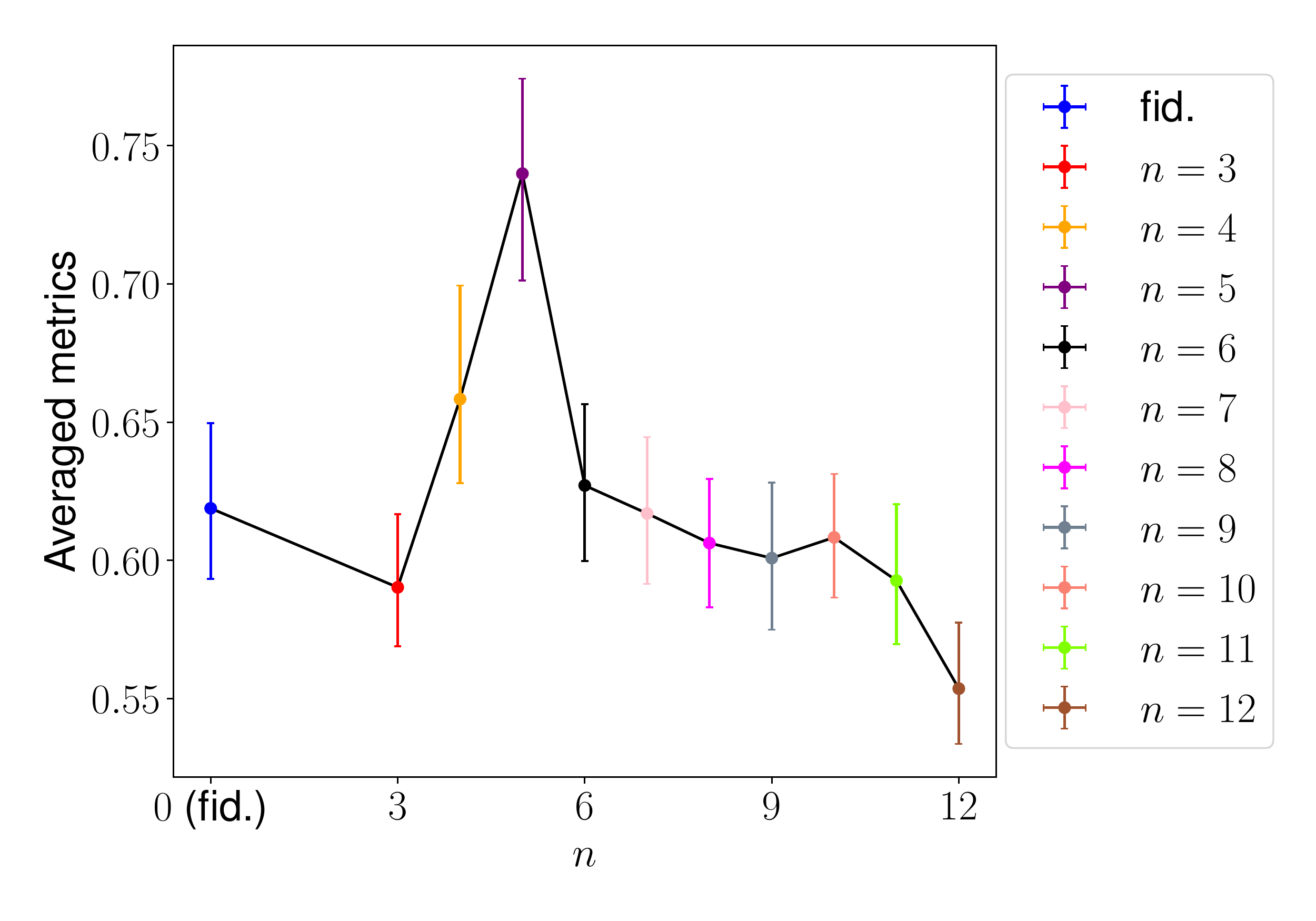}
    \end{minipage}
   
    \caption{\textit{Left panel}: On the first row a radar chart created with the metric values obtained on models optimised on the first set of constraints. The metric values for the best-fitting models are represented by stars, the median values of all MCMC samples are joined with a solid line, and the shaded areas represent $68$ per cent of the distribution. We only present the fiducial model and the ones that are the best according to a Bayesian criterion or the metrics. On the second row, a graph of the median of the averaged metrics distributions of each modelling. Errors represent $68$ per cent of the distribution. \textit{Right panel}: same radar chart and graph but for the second set of multiple images.}
    \label{fig:spider}
\end{figure*}

Radar charts shown in Fig.~\ref{fig:spider} present the different metric values, the plain lines show the median, the shaded areas $68$ per cent of the distribution and the stars represent the metric values for best-fitting models. For more clarity, we only show models that are the best according to either a Bayesian criterion or the median of the posterior distribution of the average of the six metrics obtained for a given modelling for all realisation of constraints. The model which is the best according to a metric is not necessarily showed.

The metrics improved by adding a perturbation include the 2D convergence in the constrained area, the multiple images mapping and the radial mass profiles. Indeed, the reconstruction of multiple images positions is the metric where perturbed models show the most significant enhancement. The fiducial model, on the first set, has a metric value around $2.5$ times lower than the best perturbed model ($n=7$ with a metric value of $1$). On the second set, it is only around two thirds of the best perturbed model metric value ($n=5$). Interestingly, perturbed models on the first realisation of constraints have higher RMS on multiple images used during optimisation and suffer from a starting point (i.e. the fiducial model) that have an RMS $50$ per cent higher than the one on the other set, but they performed better on this extra set of multiple images. 
 
Enhancement on the 2D mass distribution in the constrained area can be expected from results seen in Fig.~\ref{fig:err_relat}. Most notably, all perturbed models from each realisation of constraints have a metric value between $0.6$ to $0.9$. Metric values for fiducial models are around $0.7$ or lower, with the one on the second set of multiple images performing better, at the level of the worst perturbed modellings (i.e. models with $n>8$). The same behaviour happens for mass profiles when fiducial models have metric values equivalent to the ones of the worst perturbed models. Thus, the major improvements due to the perturbation are on the reconstruction of multiple images positions, and smaller ones are seen on the 2D mass distribution and the radial mass profiles.

Source positions, effective Einstein radii and mass of substructures present equivalent values between fiducial and perturbed models, with the fiducial performing more than some of the perturbed models on Einstein radii for the second set or the mass substructure on the first set of constraints. Also, the improvements on the multiple images but not on the source positions indicate that perturbed models are as biased as the fiducial model in determining positions in the source plane. We can expect from the enhancement of the 2D mass distribution that the bias is mainly a constant shift on all images which do not affect the second derivative of the lensing potential. However, our constraints are unable to break this degeneracy between different source planes and other observables such as time delay between multiple images may reduce this bias for fiducial and perturbed models. Similarly, the mass reconstruction improvements that do not include a better estimation of the substructure mass show that the enhancement is only on the cluster-scale components. Thanks to the results on the second set of constraints where refinements on the mass profiles are negligible, we can assess that the non-radial components have a better reconstruction in perturbed models with $n<8$ compared to the fiducial one.

\subsection{Bayesian criteria accuracy}
\label{sec:bayes_accuracy}

To assess which models are the best according to the metrics, for all the posterior distribution we compute the mean of the different metrics values for each modelling which are shown in Fig.~\ref{fig:spider}. They are favouring models with $n=6$ and $n=7$ almost equivalently on the first realisation of multiple images and with $n=5$ for models optimised on the second realisation. All perturbed models are better than the fiducial one for the first set of constraints. However, only the ones with $n=4,5,6,7$ are better on the second set, the other perturbed ones being equivalent or worst. Models with $n\geq8$ tend to be rejected by the metrics and they are also the models that have more parameters than the number of constraints, suggesting an over-fitting. The model with $n=3$ is only better on the first set, this is partially due to its good performance on the substructure masses (i.e. it is the best models among all others on both sets for this metric). 

Notably, there is a significant difference in modelling performances according to the averaged metrics between the results on both sets of constraints. This difference is the highest for parametric-only models with an improvement of $36$ per cent on the second set of constraints compared to the first one. In comparison, perturbed modellings show gains of less than $20$ per cent between both sets, and if we except models with $n=5$, these gains are below $12$ per cent. Hence, perturbed modellings cannot keep all enhancements provided by the parametric-only approach used to define priors on their parametric components but show less dependency on the set of constraints.

We can now assess the information given by all the criteria by comparing Fig.~\ref{fig:spider} with Fig.~\ref{fig:bayes_indicator}. BIC and AIC have selected models with $n=5$ for both set of constraints, they chose the right one for the second set and they were close for the first. We note that the AIC is little bit better as it favours equivalently models with $n=6$ and $n=5$ while the BIC rules out the former. They both have the advantage of strongly penalising models based on their number of parameters while the DIC and $-$Log$(\mathcal{E})$ show them to be equivalent. These last two criteria have a kind of Occam's razor, with an effective number of parameter for the former and the ratio of the parameter volume between the posterior and the prior distribution for the latter. Nevertheless, the effective number of parameter based on the Bayesian complexity \citep{trotta2008} is increased with $n$ but is rather almost equivalent for models with higher $n$. There is the same pattern for the evidence, as there are an increasing number of correlated or unconstrained parameters, thus more data are necessary to accurately compare model with a higher $n$. Without considering this fact we would think that the DIC and $-$Log$(\mathcal{E})$ support models with $n=12$ instead of $n=6$ for the first set of constraints and $n=8$ instead of $n=5$ for the second one, respectively. Thus, considering these minimal adjustments into account DIC and $-$Log$(\mathcal{E})$ both support model with $n=6$ on the first set and $n=5$ on the second one in agreement with the metrics. Looking at $n=n_{\rm plateau}$ when both of these criteria reach their plateau (see Section~\ref{sec:criteria}) leads to the right results. Thus, they should be preferred to the BIC and AIC, which only agree with the metric on the second set of constraints instead of both sets.

We can also look at the correlation between the averaged metric distributions and their associated distribution of $\chi^2_\nu/\nu$ (e.g. Fig.~\ref{fig:bayes_indicator}). As seen on the figure, we can not assess a strict anti-correlation between distributions on both sets of constraints. In particular, models with $n\geq8$ have lower $\chi^2_\nu/\nu$ in comparison to models with smaller $n$ that are favoured by the metrics. Hence, $\chi^2_\nu/\nu$ is not a good indicator of the model quality and can not be used alone to discriminate between different modellings. However, most models with lower $\chi^2_\nu/\nu$ values show better or equivalent performance on the averaged metrics. This trend is followed by a drop of metric values at $n=8$ and $n=6$ for the first and the second set of constraints, respectively. As these drops are disfavouring models with a higher number of free parameters, this can indicate an over-fit. Nevertheless, the limit on the number of parameters seems to depend on other properties than the number of multiples images and associated sources which are equivalent between both sets. In the light of this result and on criteria based on $\chi^2_\nu/\nu$ distributions (e.g. AIC, BIC and DIC), model quality can still be fairly assessed by combining $\chi^2_\nu/\nu$ values with a discussion on the number of free parameters as included in these criteria.

We are now able to complete our method on how we can select the best number of B-splines in a real case. Here, we have chosen the error on the multiple images positions $\sigma_{i,j}$ to be equal to $0.2$~arcsec for each image. Models should be run with a $\sigma_{i,j}$ comparable to the minimal RMS values possible on the multiple images positions to reproduce the situation obtained here, and the best modelling should be chosen according to the BIC or AIC best model or according to $n_{\rm plateau}$ values given by other criteria. A new run should be done with the selected model with a value of $\sigma_{i,j}$ comparable to the error reached in the previous run to obtain a correct estimation of model errors as it is already done in the fiducial \textit{Lenstool} method \citep{Bergamini2019,Caminha2019,Rescigno2020}.

\subsection{Evolution of the averaged metrics with the number of constraints}

To probe the correlation with the number of multiple images and the best $n$ according to the metrics, we reproduce the procedure detailed in Section~\ref{sec:simulated} to create multiply-imaged systems sets with increasing size. We start with a set of around $20$ elements that we increase by a ten until we reach approximately $110$ multiple images. We use the two sets defined previously to account for a set of $50$ elements instead of creating a new one.

\begin{figure*}
    \begin{minipage}{0.49\linewidth}
    \centering
    \includegraphics[width=1.\linewidth]{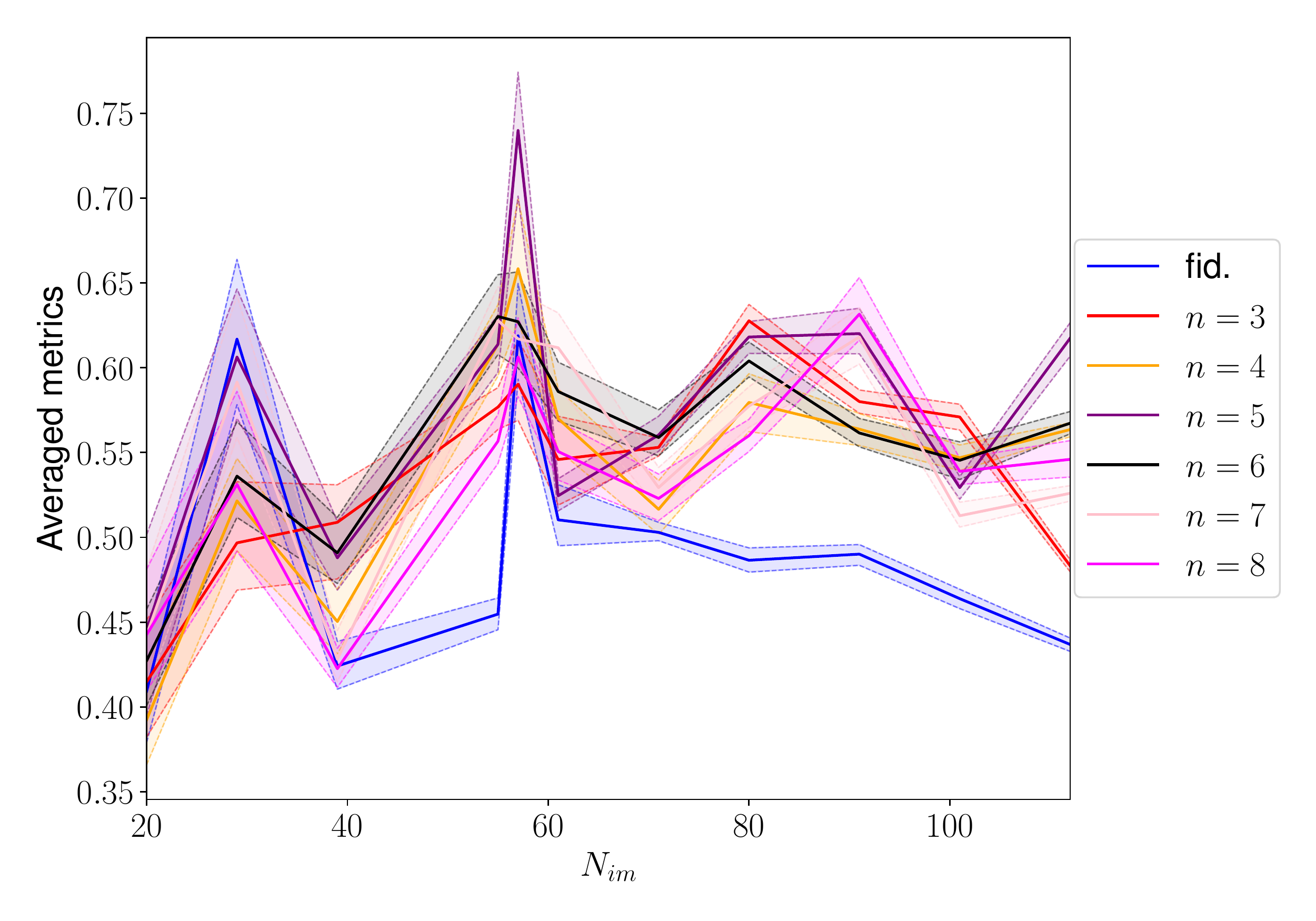}
    \end{minipage}
    \begin{minipage}{0.49\linewidth}
    \centering
    \includegraphics[width=1.\linewidth]{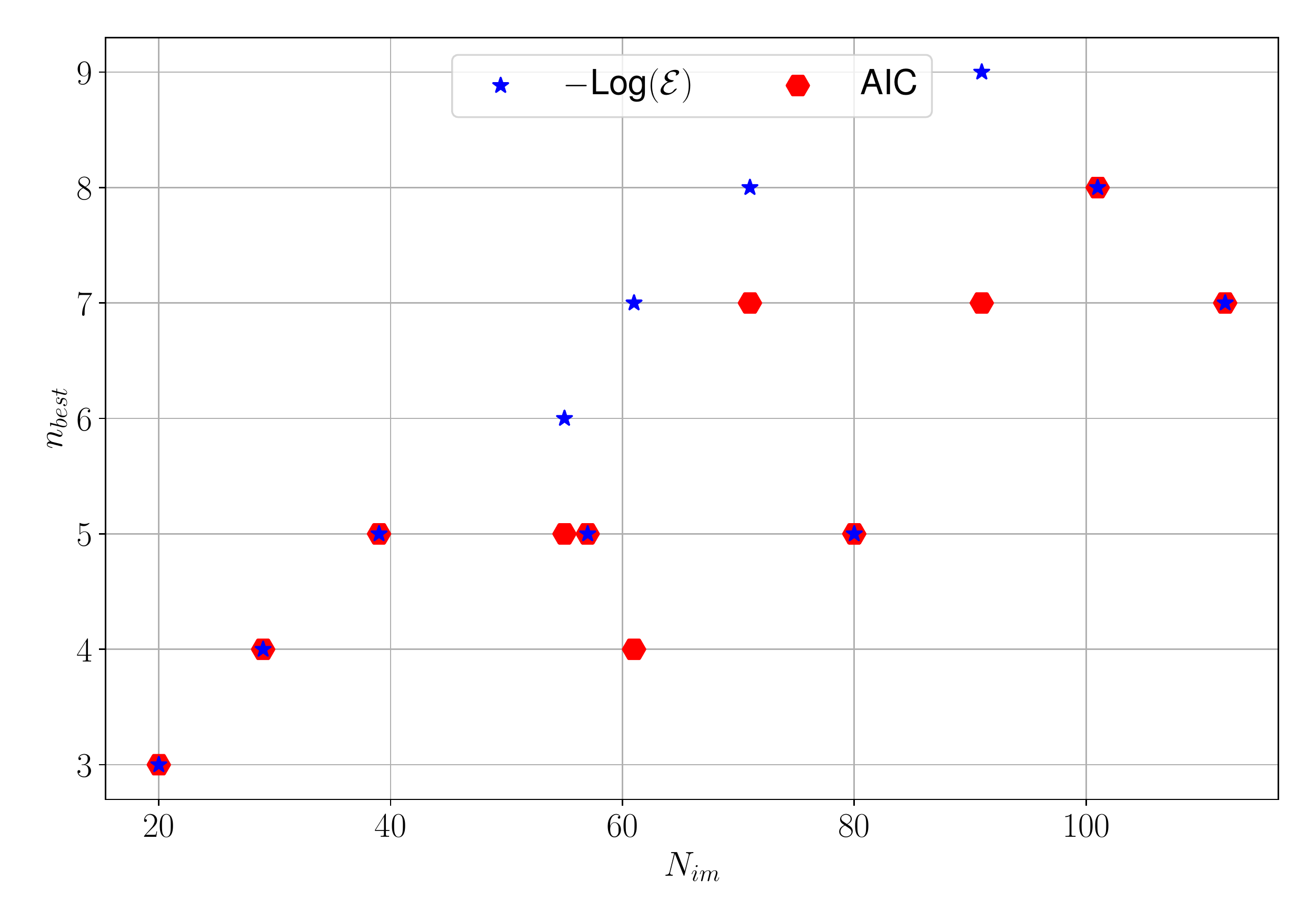}
    \end{minipage}
   
    \caption{\textit{Left panel}: Graph of the averaged metrics distribution in function of the number of multiple images used during the optimisation. The plain lines and the shaded areas represent the median and $68$ per cent of the distribution, respectively. We only present models with $n\leq8$ and fiducial ones for more clarity. \textit{Right panel}: Estimation of $n_{\rm best}$ according to the AIC and $-$Log$(\mathcal{E})$ in function of the number of multiple images. We do not show the BIC and DIC as they produce results very similar to the AIC and $-$Log$(\mathcal{E})$, respectively.}
    \label{fig:best-n}
\end{figure*}

For each of these sets, we repeat almost the same methodology of modelling and analysis that has been done in the previous section. Hence, we obtain the best $n$ according to each Bayesian criteria and the averaged metrics distributions that are presented in Fig.~\ref{fig:best-n} in function of the number of multiple images. As AIC and BIC show similar results we only present the former in Fig.~\ref{fig:best-n}. Similarly, we only represent the $-$Log$(\mathcal{E})$ to account for the two remaining criteria. The best $n$ according to the median of the averaged metric distribution is referred to as $n_{\rm best}$. As we already experienced over-fitting cases for models with a high $n$, we do not optimise models up to $n=12$ for sets with a smaller number of multiple images (i.e. sets with less than $50$ images). Following the discussion on the DIC and $-$Log$(\mathcal{E})$ in Section~\ref{sec:bayes_accuracy}, we choose the model with the lower $n$ in similar cases. 

According to all Bayesian criteria, the best modellings have $n$ values correlated with the number of constraints, especially when this number is below $80$. For a larger number of multiples images, the best $n$ does not follow the same increasing trend and seems to reach a plateau with values between $5$ and $8$. Interestingly, modellings with $n\geq8$ can not capture more information from the constraints than the ones with lower $n$, which could indicate the limit of the flexibility added to the reconstruction by the perturbative approach. The increasing complexity of the optimisation problem could also play a part in these results. However, these patterns are not fully supported by the metrics as $n_{\rm best}$ values tend to favour mostly models with $n$ between $5$ and $8$. Most models tested for sets of $30$ multiple images or less agree at $1\sigma$ according to the metrics which limits the determination of $n_{\rm best}$ in this range. Hence, if we do not take into account results on these sets, there is no clear correlation with the $n_{\rm best}$ and the number of multiple images nor with the number of constraints (i.e. the number of multiple images where the number of systems is subtracted). Other parameters related to the positions of multiply-imaged systems such as the distribution of distances between them or the systems geometries can lead to a better understanding of the $n_{\rm best}$ evolution. We searched for such correlations between these positions and the mesh of B-splines without success. 

The median of the averaged metric distribution give us more insights about the pattern followed by $n_{\rm best}$. Indeed, there is a weak increasing trend of the median values with the number of multiple images until this number reaches approximately $50$ where the median is at its peak for $n=5$. As there are no refinements for sets with a larger number of constraints, this confirms the limitation of the perturbative approach. For most sets, modellings with $n\geq8$ are unable to outperform models with a lower $n$ even with more constraints. Nevertheless, perturbed models are still ahead fiducial ones on 2D mass maps, multiple images and mass profiles metrics even on sets with less than $30$ multiple images. These enhancements are less substantial than the ones presented in Section~\ref{sec:metric}, which explains the lower performance on the averaged metrics. Hence, depending on the physical properties extracted from models, the addition of a perturbation is not always pertinent for a small set of constraints.

Finally, we use results from all these sets of multiple images to probe the consistency of the estimation of $n_{\rm best}$ by the Bayesian criteria. We use the models that have an averaged metric distribution that agrees at $1\sigma$ with the one with $n_{\rm best}$ and the mean of the favoured rank modellings for each criterion. Hence, all four criteria have similar results when estimating $n_{best}$. They favoured one of the best models (i.e. modelling in agreement within $1\sigma$ with $n_{best}$) from $6$ to $8$ out of $11$ sets; DIC and $-$Log$(\mathcal{E})$ obtain both the best performance. Considering the rank estimation, the mean among all set for each criterion are between $3.36\pm2.50$ for the DIC to $4.45\pm2.91$ for the BIC; AIC and $-$Log$(\mathcal{E})$ present result similar to the DIC. Then, we can conclude that all criteria have the same kind of performance with a small refinement provided by DIC and $-$Log$(\mathcal{E})$ compared to AIC and BIC. However, there are not consistently predicting one of the best modellings. We can assume that depending on the realisation of multiple images, models that agree at $1\sigma$ with $n_{\rm best}$ could be the best on another set. Then, these estimations, especially for the DIC and the $-$Log$(\mathcal{E})$ can be reasonably trusted even though they are not perfect.

\subsection{Computation time}
It is important to note that adding this perturbation is more time consuming compared to the parametric-only modelling. This slow down is mostly due to the increasing number of free parameters that need to be estimated. To give a comparison, fiducial models take almost 5 hours on 15 Intel Xeon E5-2640 v3 cores at 2.6GHz to be optimised. If we add a perturbation with $n=8$, this takes almost 15 hours, but this time is gradually increasing. For example, with a perturbation with $n=3$, the computational time is nearly the same as without perturbation. This scaling is due to the number of B-splines coefficient sampled, which is increasing as $n^2$. To be more specific, a single evaluation of the gradient of a B-spline surface is only two times longer than for one dPIE. Nonetheless, improvements in the implementation may speed up the modelling process in the future. In the case of already existing strong lensing models, it is possible in a very reasonable amount of time to modify them to incorporate a perturbative surface. 

\section{Summary and conclusion}
\label{sec:summary}
In this paper, we have presented a new hybrid approach for the reconstruction of the cluster cores with strong lensing constraints. It combines the \textit{parametric} modelling implemented in the software \textit{Lenstool} with \textit{free-form} components made of B-splines functions. It combines the advantages of both approaches by providing a reconstruction of cluster mass distribution that has the robustness of the first approach with the increased flexibility of the second. We tested this new method on the realistic simulation of a bi-modal galaxy cluster called \textit{Hera} which is one of the two simulated clusters chosen to compare the different modelling techniques involved for the HFF challenge \citep{meneghetti2017}. 

We found that the perturbative models showed a better reconstruction of the multiple images positions with an RMS $~3-4$ times lower than the fiducial parametric-only modelling. The estimation of the surface mass density was improved as well, with an error spread reduced in the case of perturbed models. The reproduction of the mass density radial profiles, Einstein radii and the mass in substructures were comparable in quality to the one obtained by the fiducial modellings with small improvements on the radial profiles in the constrained area. We note a dependency of these reconstructions on the realisation of constraints used during the optimisation, which is still present on perturbed models but with a smaller impact.

We quantified these improvements using an ensemble of metrics based on the previous quantities compared to the simulated data of the cluster. We also used those metrics to assess which Bayesian criteria can be used to choose the appropriate number of B-spline functions and found that they were all good candidates when considering the most favoured models (AIC and BIC) or their $n_{\rm plateau}$ (DIC and $-$Log$(\mathcal{E})$).

After demonstrating the ability of this method to reconstruct more accurately the mass distribution of a simulated cluster, the next step will be to apply it on real data such as the Hubble Frontier Field Clusters \citep{lotz2017}. The results on such clusters could also be compared to other similar techniques to pursue the effort in analysing their systematic biases \citep{Priewe2017,rainey2020,meneghetti2017}. In addition, the difference with previous models constructed with the parametric-only method of \textit{Lenstool} can give us insights on modelling choices. Especially, recent models of Abell 370 have used an external shear as a first-order perturbation, but no external structures have been identified yet to justify it \citep{lagattuta2017,lagattuta2019}. Our \textit{free-form} components could also enhance the reproduction of the multiple images positions, but unlike the external shear it would be directly linked  with the shape of the mass distribution in the cluster core. 
A better reproduction of these positions will also help the source reconstruction of multiply-imaged systems by reducing the bias due to the error on each image position. Future work may implement additional elements in the likelihood to take into account all information contained in the light distribution of each image and enhance such reconstructions. 

We make this method publicly available trough its implementation in the \textit{Lenstool} software. More information about how B-splines are computed can be found in appendix~\ref{sec:comp_Bspline} as well as an online documentation for end users\footnote{\url{https://git-cral.univ-lyon1.fr/lenstool/lenstool/-/tree/Bspline-potential}}.

Ongoing survey program such as Beyond Ultra-deep Frontier Fields and Legacy Observations (GO-15117, PIs: Steinhardt \& Jauzac, \citealt{Steinhardt2020}) will complete the current \textit{HST} data on the HFF clusters with high-resolution weak-lensing measurements of their outskirts. It has been shown that the neighbouring substructures highlighted by weak-lensing analysis can bias the reproduction of the cluster core \citep{Jauzac2016,mahler2018} through a shear effect. Thus, a combination of state-of-the-art weak lensing reconstruction such as \textit{Hybrid}-Lenstool could be combined with our free-form approach \citep{jullo2014,niemiec2020}. Such methods can model self-consistently clusters on all scales and take most of the information contained in the lensing measurements to produce more accurate mass distribution for cluster lenses. It is especially crucial as they allow us to probe the dark matter properties and the cluster physics, but also enable a detailed study of the high-redshift Universe thanks to the magnification they provide.

\section*{Data availability}
The data underlying this article will be shared on reasonable request to the corresponding author.
\section*{Acknowledgements}

BB, BC and JR acknowledge support from the ERC starting grant 336736 (CALENDS). The authors are grateful to the referee for careful reading of the manuscript and valuable suggestions and comments which helped to improve it.




\bibliographystyle{mnras}
\bibliography{biblio} 



\appendix

\section{Computing the B-spline surface}
\label{sec:comp_Bspline}

To compute the B-spline surface at Eq.~\ref{eq:pert_surf}, we used the De Boor algorithm which is written in pseudo-code in Table~\ref{tab:de_boor_alg}. This algorithm computes the following sum:
\begin{equation}
    \sum^m_{j=1} C_{j} B_{j,p,t}(u)
\end{equation}
Where $p$ is the polynomial degree, $C$ the coefficient vector, $t$ the knot vector, $m$ the number of basis functions and $u$ the point where the B-spline is evaluated. Before applying the algorithm, the knot index $k_u$ associated with $u$ has to be found such that $t_{k_u}\leq u < t_{k_u+1}$. To compute the surface, the algorithm has to be applied to compute two different sums, each associated with one axis. The first one is:
\begin{equation}
    \sum^m_{l=1} C_{j,l} B_{l,p,t_x}(x)
\end{equation}
The coefficient $C_{j,l}$ are substituted by the $C_j=\sum^m_{l=1} C_{j,l} B_{l,p,t_x}(x)$ which allows to apply the algorithm a second time to compute an evaluation of the surface:
\begin{equation}
    \sum^m_{j=1} C_{j} B_{j,p,t_y}(y)
\end{equation}
All $C_j$ do not have to be computed as we only need the ones with $j=k_y-p,...,k_y$ to apply the algorithm where $k_y$ is the knot index associated with $y$ and $t_y$. The surface is then multiplied by $D_{ls}/D_s$ to obtain the perturbation added to the lensing potential.

Most of the lensing quantities can be computed with some linear combinations of the lensing potential derivatives. Hence, we will just define how to compute the derivative of the B-spline surface associated with the potential. As derivatives of a B-spline surface are also B-spline surfaces with different coefficients, knot vectors and polynomial degrees, we will only express that change since the computing procedure of these surfaces is the same as before. We will only consider the derivation along the X-axis as it can be generalised to the Y-axis. The coefficient $C^x_{j,l}$ associated with this derivative are expressed as:
\begin{equation}
    C^x_{j,l}=\frac{p}{t_{x,j+p+1}-t_{x,j+1}}(C_{j+1,l}-C_{j,l})
\end{equation}
Substituting $t_x$ by $t_y$ and $j$ to $l$ gives the expression for the Y-axis. We note that there are $m-1$ basis functions on the considered axis instead of $m$, with a polynomial degree of $p-1$ and $p$ on the X-axis and Y-axis, respectively. The knot vector associated with the derivative is also modified, as its first and last terms are removed. Derivative of higher order are computed by applying recursively the preceding procedure.

\begin{table}
    \centering
	\begin{tabular}{ll} 
	\hline\hline
     \textbf{De Boor Algorithm}\\
    \hline\hline
    \textbf{Input} $p$: Polynomial degree\\
    \hspace{0.75cm}$k_u$: knot index\\
    \hspace{0.75cm}$u$: evaluation point\\
    \hspace{0.75cm}$\left(t_j\right)_{j\in[0,N-1]}$: knot vector\\
    \hspace{0.75cm}$\left(C_j\right)_{j\in[0,m-1]}$: coefficient vector\\
    \textbf{Variable} $(d_j)_{j\in[0,p-1]}$\\
    \textbf{Output} $d_p$: Surface evaluation \\
    \hline
    \textbf{For} \textit{j}$=0,..,p$, \textbf{Do}\\
    \hspace{1cm}$d_j\longleftarrow C_{j+k-p}$\\
    \textbf{End For}\\
    \textbf{For} \textit{r}$=1,..,p$, \textbf{Do}\\
    \hspace{1cm}\textbf{For} \textit{j}$=p,..,r-1$, \textbf{Do}\\
    \hspace{2cm}$\delta\longleftarrow(u-t_{j+k-p})/(t_{j+k+1-r}-t_{j+k-p})$\\
    \hspace{2cm}$d_j\longleftarrow(1-\delta)d_{j-1}+\delta d_{j}$\\
    \hspace{1cm}\textbf{End For}\\
    \textbf{End For}\\
    \textbf{Return} $d_p$\\
    \hline\hline
	\end{tabular}
	\caption{Implementation of the De Boor algorithm used in the \textit{Lenstool} software.}
	\label{tab:de_boor_alg}
\end{table}


\bsp	
\label{lastpage}
\end{document}